\documentclass[journal]{IEEEtran}
\pdfoutput=1 
\usepackage{algorithm}
\usepackage{algorithmic}
\usepackage{bbm}
\usepackage{amsmath}
\usepackage{multirow}

\usepackage{graphicx}
\usepackage{epstopdf}
\usepackage{extarrows}
\usepackage{float}
\usepackage[latin1]{inputenc}
\usepackage{tikz}
\usepackage{mathrsfs}
\usepackage{amssymb}
\usepackage{stmaryrd}
\usepackage{mathabx}
\usepackage[normalem]{ulem}
\usepackage{flushend}
\usepackage{amsfonts}
\usetikzlibrary{trees,arrows,automata, positioning, shapes}
\usepackage{tikz-qtree}
\usepackage{multirow}
\hfuzz=\maxdimen
 \usepackage{indentfirst}
\usepackage{subfigure}
\usepackage{array}
\newcolumntype{C}[1]{>{\centering\arraybackslash}p{#1}}
\usepackage[pdftex, bookmarksnumbered, bookmarksopen, colorlinks, citecolor=blue, linkcolor=blue]{hyperref}
\usepackage{pifont}

\newtheorem{definition}{Definition}
\newtheorem{theorem}{Theorem}
\newtheorem{lemma}{Lemma}

\newtheorem{remark}{Remark}
\newtheorem{example}{Example}
\newtheorem{proposition}{Proposition}

\begin{document}

\title{Opacity of Parametric  Discrete Event Systems: Models, Decidability, and Algorithms}

\author{
    Weilin Deng, Daowen Qiu$^{\star}$, and Jingkai Yang
    \thanks{
    Weilin Deng is with the School of Internet Finance and Information Engineering, Guangdong University of Finance, Guangzhou, 510521, China (e-mail: williamten@163.com).}
    \thanks{Daowen Qiu (Corresponding author) is with the Institute of Quantum Computing
and Computer Theory, School of Computer Science and Engineering, Sun Yat-Sen University, Guangzhou, 510006, China
(e-mail: issqdw@mail.sysu.edu.cn).}
    \thanks{Jingkai Yang is with the School of Mathematics and Statistics, Yulin Normal University,
    Yulin, 537000, China (e-mail: yangjingkai711@163.com).}
}

\maketitle

\begin{abstract}
Finite automata (FAs) model is a popular tool to characterize discrete event systems (DESs)
due to its succinctness.
However, for some complex systems,
it is difficult to describe the necessary details by means of FAs model.
In this paper, we consider a kind of extended finite automata (EFAs) in which each transition carries a predicate over  state and event parameters.
We also consider a type of simplified EFAs, called Event-Parameters EFAs (EP-EFAs),
where the state parameters are removed.
Based upon these two parametric  models,
we investigate the problem of opacity analysis for parametric DESs.
First of all, it is shown that EFAs model is more expressive than EP-EFAs model.
Secondly, it is proved that the opacity properties for EFAs
 are undecidable in general.
Moreover,  the decidable opacity properties for EP-EFAs are investigated.
 We present the verification algorithms for current-state opacity, initial-state opacity and infinite-step opacity,
 and then discuss the complexity.
 This paper establishes a preliminary theory for the opacity of parametric DESs,
 which lays a foundation for the opacity analysis of complex systems.
\end{abstract}

\begin{IEEEkeywords}
Opacity, discrete-event systems, parametric  finite automata, extended finite automata
\end{IEEEkeywords}

\section{Introduction}

Over the last ten years, the problem of opacity analysis for discrete event systems (DESs) received considerable attention.
\emph{Opacity} is an important security property, which was initially introduced in computer science to
analyze cryptographic protocols.
Roughly speaking, a DES is said to be opaque, if the intruder cannot determine the occurrence of the secret behavior by his observations to the system.
Finite automata (FAs) model is a popular tool to describe DESs in logical level due to its succinctness \cite{desbook}.
The notions of language-based opacity  \cite{opacity-review}, \cite{l-opacity},
current-state opacity \cite{cso}, initial-state opacity \cite{iso,ifo},
infinite-step opacity \cite{infinite} and pre-opacity \cite{pre-opacity}
for FAs model were well investigated in recent years.
In addition, the opacity enforcement based on the techniques of supervisory control and output obfuscation
were proposed (e.g., see \cite{supervisory-enforcement1}-\cite{output-enforcement3} and references therein).

\par
For some complex  systems,
it is difficult to describe the necessary details and analyze their opacity properties by means of FAs model,
and thus some extended models are necessary.
Actually, the opacity properties for various extended models were investigated recently,
such as time systems \cite{timed-opacity,timed-opacity2}, networked systems \cite{network-opacity1,network-opacity2},
Petri nets \cite{petri-opacity1}-\cite{petri-opacity3},
cyber-physical systems \cite{opacity-cps1,opacity-cps2},
probabilistic systems \cite{probilistic-opacity1, probilistic-opacity2}
 fuzzy systems \cite{fuzzy-opacity1, fuzzy-opacity2}, and the other systems \cite{efa1}-\cite{efa3}.

\par
In the field of system modeling, \emph{control flow} refers to the possible sequences of the interactions between a system and its environment, and \emph{data flow} refers to the constraints on data parameters in the interactions \cite{learning}.
FAs model well describes control flow, but fails to capture data flow and
the mutual influence between control flow and data flow efficiently.
A typical example is modeling network protocols, where the models must characterize how different parameter values in sequence numbers, user IDs, socket IDs, etc., affect the control flow.
Another easy-to-understand  example is modeling the process of web-site registering that usually requires
a user to provide her/his \emph{identical} password twice
(see Examples \ref{ex:1}-\ref{ex:2} and Remark \ref{remark:1} in Section II for details).
Obviously, it is difficult and inefficient for FAs model to do such things.

\par
To address this problem, in this paper, we also consider a kind of extended finite automata (EFAs), in which the  states and  events are both augmented with parameters, and each transition carries a predicate and an update function over these parameters.
 The EFAs model is a powerful but complicated tool.
 It is hard to analyze some properties of EFAs, and we prove that the opacity properties of EFAs are undecidable.
Thus, we also consider a simplified EFAs model, called Event-Parameters EFAs (EP-EFAs),
where the state parameters are removed.
By means of the transitions carrying predicates over parameters,
the models of EFAs and EP-EFAs improve FAs model in efficiently representing and handling some complex systems where
control flow, data flow and the interactions between them are required to be characterized.
\par
In general, EFAs and EP-EFAs can be viewed as a special type of infinite and finite state models, respectively,
with infinite alphabet, which have been well investigated in computer science (e.g., see \cite{esfa}-\cite{infinite-alphabet3}).
For the general infinite state automata (ISA), only a few properties are decidable \cite{esfa}, \cite{sft},
However, for some types of ISA, there exist quite a few decidable properties, e.g.,
 the properties of reachability, simulation and eventuality of \emph{Well-Structured ISA}
 are all decidable \cite{infinite-alphabet1}.
  On the other hand, for the finite state models with infinite alphabet,
  there are many decidable properties, as well as undecidable  properties \cite{infinite-alphabet2}.
 For example, the emptiness and language inclusion of \emph{1N-RAs} are decidable;
 however,  its universality and equivalence are undecidable \cite{infinite-alphabet3}.

\par
In this paper, the aforementioned EFAs and EP-EFAs are referred to as parametric DESs collectively.
 We would like to establish a preliminary  theory for the opacity of parametric DESs,
 which lays a foundation to analyze the opacity of some complex systems.
 To the best of our knowledge, this is the first study on the opacity analysis of parametric DESs.
 The main contributions of this paper are as follows.
 \par

\begin{enumerate}
    \item
     Two parametric models, i.e., EFAs and EP-EFAs, are introduced for DESs,
     and then it is proved that the latter can be simulated by the former
     but the reverse does not hold.
     This means that EFAs model is more expressive than EP-EFAs model.
     We also illustrate that these two parametric models are both more expressive and efficient than FAs model.
  \item
  We formulate  the current-state opacity, initial-state opacity and infinite-step opacity  for parametric DESs,
   and  then prove that these opacity properties for \emph{EFAs} are all undecidable in general.
   The basic idea of the proof is reducing the halting problem of Two-Counter Machines (2CMs)
   to the verification of the opacity properties.

  \item
  We investigate the decidable opacity properties for  \emph{EP-EFAs}.
   Based on the \emph{symbolic observer}, the verification algorithms for current-state opacity, initial-state opacity and infinite-step opacity  are provided, and the complexity is analyzed.

\end{enumerate}

\par
The rest of this paper is organized as follows.
The system models for parametric DESs are introduced and investigated in Section II.
The problem formulation and necessary assumptions are provided in Section III.
In Section IV, the opacity properties of  EFAs are proved to be undecidable, and
in Section V,  the decidable opacity properties of  EP-EFAs are studied.
Finally, Section VI concludes this paper.

\section{Parametric  Models}
In this section, we present some notations, and introduce two parametric  models: extended finite automata (EFAs) and
  Event-Parameters EFAs (EP-EFAs),
  and then discuss their expressiveness and efficiency.

\par
Let  $\mathbb{N}$ be the set of natural numbers,
 and $[m:n]$ be the set of integers $\{m,m+1,\ldots,n\}$.
 Let $\Sigma$ be an alphabet,
  $\Sigma^{*}$ be the set of finite strings over $\Sigma$ including  empty string $\epsilon$,
  $\Sigma^{k}$ be the set of strings of length $k$,
and  $\Sigma^{\leq k}$ be the set of strings of length $i$, $i \in [0:k]$, over $\Sigma$.
A \emph{language} $L$ over $\Sigma$ is a subset of $\Sigma^{*} $.
We denote by $|\Omega|$  the number of elements in the set of $\Omega$,
and by $|s|$ the length of the string $s \in L$ with a slight abuse of notation.
\par

A \emph{discrete event system (DES)} is usually modeled as a \emph{finite automaton}
$H=(Q, q_{0},\Sigma, \delta)$ \cite{desbook}, where $Q$ is the finite set of states,
$Q_{0} \subseteq Q$ is the set of initial states,
$\Sigma$ is the finite set of events,
and $\delta:Q \times \Sigma \rightarrow Q$ is the deterministic (partial) transition function.
The transition function $\delta$ can be extended to domains $Q \times \Sigma^{*}$ and $2^Q \times 2^{\Sigma^{*}}$
by the usual manner.
The \emph{generated language} by $H$ is
$L(H) = \{s | \exists q_{0} \in Q_{0}, q \in Q, \text{ s.t. } q = \delta(q_{0},s) \}$.

\par
A Boolean algebra is a tuple $\mathcal{A}=(\mathcal{U}, \Psi, \llbracket \bullet \rrbracket)$,    
where $\mathcal{U}$ is the universe of discourse,
 and $\Psi$ is the set of  predicates closed under the Boolean connectives, substitution, equality and if-then-else terms \cite{esfa}.
 The element $\varphi \in \Psi$ is called an $\mathcal{U}$\emph{-predicate} in $\mathcal{A}$, or just \emph{predicate} when
 $\mathcal{U}$ and $\mathcal{A}$ are clear from the context.
The denotation function $ \llbracket \bullet \rrbracket: \Psi \rightarrow 2^{\mathcal{U}}$
maps a predicate to the valuations of variables that make the predicate true.
 Hence, for any $\varphi, \psi \in \Psi$,
$\llbracket \varphi \wedge \psi \rrbracket =  \llbracket \varphi \rrbracket \cap \llbracket \psi \rrbracket $,
$\llbracket \varphi \vee \psi \rrbracket =  \llbracket \varphi \rrbracket \cup \llbracket \psi \rrbracket $,
$\llbracket \neg \varphi \rrbracket = \mathcal{U} \backslash \llbracket  \varphi \rrbracket $ \cite{solver2}.
For the \emph{true predicate} $\top$ and \emph{false predicate} $\bot$, we have
$\llbracket \top \rrbracket = \mathcal{U}$ and $\llbracket \bot \rrbracket = \varnothing$.
For any $\varphi \in \Psi$, $\varphi$ is said to be \emph{satisfiable}, denoted by $isSat(\varphi)$,
if $\llbracket \varphi \rrbracket \neq \emptyset$.
This paper solely focuses on Boolean algebras in which the predicate satisfiability is decidable.
\par

Throughout this paper, we denote by $X$ and $Y$ the (infinite or finite) domains of event and state parameters, respectively, and denote by $x$ and $y$ (with superscript and subscript usually) the event and state parameters, respectively.
In addition, we use $a, b$ (with superscript and subscript usually) to denote the specific values of event and state parameters, respectively.

\par
The model of extended finite automata (EFAs) is defined as follows.
\par

\begin{definition} \label{def:efa}
  An \emph{extended finite automaton (EFA)} is defined as $ E = (Q, \Sigma, X, Q_{0}, Q_{m}, Y, Y_{0}, R )$ where
    $Q$ is the finite set of state tags,
    $\Sigma$ is the finite set of event tags,
    $X$ is the domain of one event parameter,
    $Q_{0} \subseteq Q$ and $Q_{m} \subseteq Q$ are the sets of tags of initial and marked states, respectively,
    $Y$ is the domain of the state parameter and
   $Y_{0} \subseteq Y$ is the domain of parameter for initial states,
   and $R$ is the set of symbolic transitions and each symbolic transition $r \in R$  is of form
   $  q \xrightarrow[k]{\sigma:\varphi:\xi} \widehat{q}$ where
  \begin{itemize}
    \item
    $q \in Q$ and $\widehat{q} \in Q$ are the tags of source and target states, respectively, which carry state parameters $y_{q} \in Y$ and $y_{\widehat{q}} \in Y$, respectively;

    \item
    $k \geq 0$, the \emph{step-length} of the transition, is the size of the tuple of event parameters in this transition;

    \item
    $\sigma \in \Sigma$ is the tag of the event, and if $k \geq 1$, it carries a $k$-tuple of event parameters
   $\langle x_{\sigma}^{1}, x_{\sigma}^{2}, \ldots, x_{\sigma}^{k}\rangle$, $x_{\sigma}^{i} \in X$, $i \in [1:k]$,
   otherwise, it  carries no event parameter;

  \item
    $\varphi$
    is the guard of transition $r$, and it is a $(Y \times X^{k})$-predicate if $k \geq 1$ otherwise a $Y$-predicate,
     and if event $\sigma$ occurs at state $q$ with the proper values of parameters to enable $\varphi$,
     then the transition $r$ may be fired;

    \item
    $\xi$, a $Y \times X^{k} \rightarrow Y$ function if $k \geq 1$ otherwise a $Y \rightarrow Y$ function,
    is responsible for updating the parameter of target state
    according to the given parameters of source state and event when the transition $r$ is fired.
     We denote by $\Xi$ the special updating function
     that does nothing.  
  \end{itemize}
If there are multiple transitions that can be fired at a state, then only one of them is fired nondeterministically.
$E$ is said to be \emph{deterministic}, if no more than one transition can be fired synchronously at each state,
i.e., for  two different  transitions
 $q \xrightarrow[k]{\sigma:\varphi_{1}:\xi_{1}} \widehat{q}_{1}$ and
$ q \xrightarrow[k]{\sigma:\varphi_{2}:\xi_{2}} \widehat{q}_{2}$,
  $\varphi_{1}(b, \langle a_{1},  a_{2}, \ldots,  a_{k} \rangle) \wedge \varphi_{2}(b, \langle a_{1},  a_{2}, \ldots,  a_{k} \rangle)$ dose not hold for any state parameter value $b$ and event parameters values
 $\langle a_{1},  a_{2}, \ldots, a_{k} \rangle$.
Moreover, the implicit $\epsilon$-selfloop $q \xrightarrow{\epsilon} q$ can be viewed as the special
  \emph{0-step-length} transition $q \xrightarrow[0]{\epsilon:\top:\Xi} q$.
 The \emph{step-length} of $E$ is defined as the maximum of the step-lengths of the symbolic transitions in $E$.
\end{definition}

\par
  Actually, the \emph{symbolic transition}  $ q \xrightarrow[k]{\sigma:\varphi:\xi} \widehat{q}$, $k \geq 1$,
  defines the set of \emph{concrete transitions}
\begin{multline}
   \Big \{(q,b)\xrightarrow{\sigma\langle a_{1}, a_{2}, \ldots, a_{k}\rangle} (\widehat{q},\widehat{b}) \Big |
(b, \langle a_{1}, a_{2}, \ldots, a_{k}\rangle) \in \llbracket \varphi \rrbracket \\
 \wedge  \widehat{b} = \xi(b,\langle a_{1}, a_{2}, \ldots, a_{k}\rangle) \Big \},
\end{multline}
where $(q,b)$ and $(\widehat{q},\widehat{b})$ are the source and target states, respectively,  and
$\sigma\langle a_{1}, a_{2}, \ldots, a_{k}\rangle$
is the \emph{parameterized event} of the concrete transition.
For example, suppose $X=Y=\{0,$ $1,2\}$, the symbolic transition
$q \xrightarrow[1]{\sigma:y_{q} = x_{\sigma}^{1}:y_{\widehat{q}} \leftarrow y_{q} + x_{\sigma}^{1}} \widehat{q}$ denotes the
set of concrete transitions
$\{(q,0)\xrightarrow{\sigma\langle0\rangle} (\widehat{q},0)$, $(q,1)\xrightarrow{\sigma\langle1\rangle} (\widehat{q},2)\}$.

  \par
  The symbolic transitions allow EFAs model to efficiently characterize the control flow (i.e., the possible sequences of fired transitions), data flow (i.e., the constraints on event parameters) and their interactions in a system.

\par

A \emph{parameterized string} is a sequence of parameterized events, and
a \emph{parameterized language} is a set of parameterized strings.
For a parameterized string
 $u = v_{1}v_{2}\ldots v_{n}$, where
 $v_{i}=\sigma_{i}\langle a_{{i}}^{1}, a_{{i}}^{2}, \ldots, a_{{i}}^{k_{i}}\rangle$, $k_{i} \geq 0$
 \footnote{ if $k_{i}= 0$, then $v_{i}=\sigma_{i}$ and
 $\mathfrak{D}(u) = \mathfrak{D}(v_{1}\ldots v_{i-1}v_{i+1}\ldots v_{n})$.},
 the \emph{data string} of $u$, denoted by $\mathfrak{D}(u)$, is obtained by stripping all the event tags $\sigma_{i}$, i.e.,
 $\mathfrak{D}(u)  = \langle a_{{1}}^{1}, a_{{1}}^{2}, \ldots, a_{{1}}^{k_{1}}\rangle$
 $\langle a_{{2}}^{1}, a_{{2}}^{2}, \ldots, a_{{2}}^{k_{2}}\rangle $ $ \ldots$
  $\langle a_{{n}}^{1}, a_{{n}}^{2}, \ldots, a_{{n}}^{k_{n}}\rangle$ is a sequence of \emph{event parameter tuples}.
 Intuitively, a data string is a sequence of data exchanges
 between the system and its environment that meet the data constraints.
   The \emph{flat data string} is obtained by flattening the parameters of a data string in order, i.e.,
   $\mathfrak{fD}(u) = a_{{1}}^{1} a_{{1}}^{2} \ldots a_{{1}}^{k_{1}}$
 $a_{{2}}^{1} a_{{2}}^{2} \ldots a_{{2}}^{k_{2}} \ldots$
  $a_{{n}}^{1} a_{{n}}^{2} \ldots a_{{n}}^{k_{n}}$.

\par

\begin{definition} \label{def:longtransition1}
  Given an EFA $E = (Q, \Sigma, X, Q_{0}, Q_{m}, Y, Y_{0}$, $R )$.
  If there exist a series of concrete transitions
  $(q_{i},b_{i}) \xrightarrow{v_{i}} (q_{i+1},  b_{i+1})$, where these
  $v_{i} $ are parameterized events,
  $i \in [1:n]$, $n \geq 1$,
  we define the \emph{combined  concrete transition} as the path
  $(q_{1},b_{1}) \xrightarrow{u} (q_{n+1},  b_{n+1})$ where
$u = v_{1}v_{2}   \ldots  v_{n}$.
\end{definition}

\par
The language between a set of source states $Q_{1} \subseteq Q$ and
 a set of target states $Q_{2} \subseteq Q$ of the EFA $E = (Q, \Sigma, X, $ $Q_{0}, Q_{m}, Y, Y_{0}, R )$
 is defined as follows.
  \begin{multline} \label{eq:efalanguage}
  L_{Q_{1}}^{Q_{2}}(E)= \big \{ u \big | \exists   q_{1} \in Q_{1}, q_{2} \in Q_{2}, b_{1}, b_{2} \in Y, \\
  \text{ s.t. } (q_{1},b_{{1}}) \xrightarrow{u} (q_{2}, b_{{2}})
  \wedge ( q_{1} \in Q_{0} \Rightarrow b_{{1}} \in Y_{0}) \big \}.
  \end{multline}
The \emph{generated language} and \emph{marked language} by the EFA $E$ are, respectively, defined as
\begin{align}
  L(E) = L_{Q_{0}}^{Q}(E)   \text{ and }   L_{m}(E) = L_{Q_{0}}^{Q_{m}}(E).
\end{align}
The \emph{data language} and \emph{marked data language} of the EFA $E$ are, respectively, defined as
\begin{equation}
 L_{d}(E) = \bigcup_{u \in L(E)} \mathfrak{D}(u) \text{ and } L_{md}(E) = \bigcup_{u \in L_{m}(E)} \mathfrak{D}(u).
\end{equation}
The \emph{flat data language} and \emph{flat marked data language} of the EFA $E$ are, respectively, defined as
\begin{equation}
L_{fd}(E) = \bigcup_{u \in L(E)} \mathfrak{fD}(u) \text{ and }
L_{fmd}(E) = \bigcup_{u \in L_{m}(E)} \mathfrak{fD}(u).
\end{equation}

\begin{figure}[h]
\centering
    \resizebox{0.44\textwidth}{!}{
             \begin{tikzpicture}[font=\large,->,>=stealth',shorten >=1pt,auto,node distance=6.5cm,semithick,
                                          every state/.style={font=\large}]
                                \node[initial, state] (q0) {$ q_{0} $};
                                \node[state] [node distance=2cm] [above of = q0] (q1) {$q_{1} $};
                                \node[state] [right of = q1] (q2) {$q_{2} $};
                                \node[accepting, state] [node distance=2cm] [below of = q2] (q3) {$q_{3} $};

                            \draw[every node/.style={font=\Large}]
                                   (q0) [right]  edge  node{$\frac{\sigma_{1}:\top:\Xi}{1}$} (q1)
                                   (q1) [above] edge  node{$\frac{\sigma_{2}:\top:y_{q_{2}}\leftarrow x_{\sigma2}^{1}}{1}$} (q2)
                                   (q2) [right] edge  node{$\frac{\sigma_{3}:y_{q_{2}}=x_{\sigma_{3}}^{1}:\Xi}{1}$} (q3)
                                   (q2)  [bend left, above left] edge node{$\frac{\sigma_{3}:y_{q_{2}} \neq x_{\sigma_{3}}^{1}:\Xi}{1}$} (q0);
                               \end{tikzpicture}
                    }
\caption{The EFA $E$  simulating the process of user registering in a web-site, where
$y_{q_{2}}$ is the parameter of state $q_{2}$, and $x_{\sigma_{2}}^{1}$ and $x_{\sigma_{3}}^{1}$
are the parameters of event $\sigma_{2}$ and $\sigma_{3}$, respectively.}
\label{fig-1}
\end{figure}
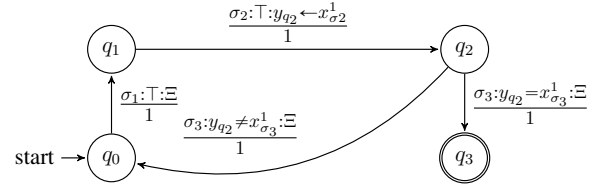

\begin{figure}[h]
\centering
    \resizebox{0.25\textwidth}{!}{%
             \begin{tikzpicture}[font=\Large,->,>=stealth',shorten >=1pt,auto,node distance=7cm,semithick,
                                          every state/.style={font=\Large}]
                                \node[initial, state] (q0) {$ q_{0} $};
                                \node[accepting, state] [node distance=4.5cm] [right of = q0] (q1) {$q_{1} $};

                            \draw[every node/.style={font=\LARGE}]
                                   (q0) [above]  edge  node{$\frac{\sigma: x_{\sigma}^{2} = x_{\sigma}^{3}}{3}$} (q1)
                                   (q0) [loop above] edge  node{$\frac{\sigma: x_{\sigma}^{2} \neq x_{\sigma}^{3}}{3}$} ();
                               \end{tikzpicture}
                    }
\caption{The EP-EFA $S$ simulating the process of user registering in a web-site
where the second and third event parameters $x_{\sigma}^{2}$ and $x_{\sigma}^{3}$ denote the password provided by user at the first and second times, respectively. }
\label{fig-2}
\end{figure}
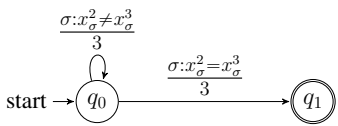

\begin{example}\label{ex:1}
The EFA $E$ shown in Fig. \ref{fig-2} simulates the process of user registering in a web-site,
where the user is required to provide his password twice for confirming its correctness. Suppose the charset for nickname
and password are both $\Omega$, then the domains of state and event parameters are $Y=X=\Omega^{*}$.
In the symbolic transition from $q_{0}$ to $q_{1}$,
  the user inputs his nickname,
  and the guard $\top$ does not block any input and the updating  function $\Xi$  does nothing.
In the symbolic transition from $q_{1}$ to $q_{2}$,
the user inputs his password for the first time (denoted by  $x_{\sigma_{2}}^{1}$),
and the updating function $\xi$ is defined as $y_{q_{2}} \leftarrow x_{\sigma_{2}}^{1}$ that means the password is stored to the target state $q_{2}$ as its parameter $y_{q_{2}}$.
In the symbolic transitions from $q_2$ to $q_3$ and from $q_2$ to $q_0$,
the user provides his password for the second time (denoted by  $x_{\sigma_{3}}^{1}$).
 If these two passwords are identical (i.e., $y_{q_{2}}=x_{\sigma_{3}}^{1}$),
then the former transition is fired, and the process goes to the final state $q_{3}$ and terminates successfully,
otherwise (i.e., $y_{q_{2}} \neq x_{\sigma_{3}}^{1}$) the latter transition is fired,
and the process fails and goes back to the initial state $q_{0}$.
\end{example}
\par
Note that the EFA $E$ shown in Fig. \ref{fig-1} is of 1-step-length.
A more concise 3-step-length EFA with only two states and two symbolic transitions
can also describe the same process.
Before introducing this, we present a simplified EFAs model that has no state parameter.

\par
\begin{definition} \label{def:EP-EFA}
  An \emph{Event-Parameters EFA (EP-EFA)} is defined as
  $ S = (Q, \Sigma, X, Q_{0}, Q_{m}, T )$,
  where
      $Q$ is the finite set of states,
    $\Sigma$ is the finite set of event tags,
    $X$ is the domain of one event parameter,
    $Q_{0} \subseteq Q$ and $Q_{m} \subseteq Q$ are the sets of initial and marked states, respectively,
   and $T$ is the set of symbolic transitions and each  symbolic transition $t \in T$ is of form
   $ q \xrightarrow[k]{\sigma:\varphi} \widehat{q}$ where
  \begin{itemize}
   \item
    $q \in Q$ and $\widehat{q} \in Q$ are the source and target states, respectively;

    \item
    $k \geq 0$, the \emph{step-length} of the transition,
    is the size of the tuple of event parameters in this transition;

 \item
    $\sigma \in \Sigma$ is the tag of the event, and if $k \geq 1$, it carries a $k$-tuple of event parameters
   $\langle x_{\sigma}^{1}, x_{\sigma}^{2}, \ldots, x_{\sigma}^{k}\rangle$, $x_{\sigma}^{i} \in X$, $i \in [1:k]$,
   otherwise it carries no event parameter;

  \item
    $\varphi$
    is the guard of transition $t$, and it is an $ X^{k}$-predicate if $k \geq 1$ otherwise the true predicate $\top$,
   and if event $\sigma$ occurs at state $q$ with the proper values of parameters to enable $\varphi$,
     then the transition $t$ may be fired.
  \end{itemize}
  If there are multiple transitions that can be fired at a state, then only one of them is fired nondeterministically.
$E$ is said to be \emph{deterministic}, if no more than one transition can be fired synchronously at each state,
i.e., for  two different  transitions
 $q \xrightarrow[k]{\sigma:\varphi_{1}} \widehat{q}_{1}$ and
$ q \xrightarrow[k]{\sigma:\varphi_{2}} \widehat{q}_{2}$,
  $\varphi_{1}(\langle a_{1}, a_{2}, \ldots,  a_{k}
 \rangle) \wedge \varphi_{2}(\langle a_{1}, a_{2}, \ldots,  a_{k}
 \rangle)$ does not hold for any event parameters values $\langle a_{1},
 a_{2}, \ldots, a_{k} \rangle$\footnote{If $k = 0$, then $\varphi_{1} = \varphi_{2} = \top$
  by the definition of symbolic
 transition. In this case, the determinism requires that
 there do not exist such two \emph{different} transitions
 $q \xrightarrow[0]{\sigma:\top} \widehat{q}_{1}$ and
 $q \xrightarrow[0]{\sigma:\top} \widehat{q}_{2}$, which is actually the condition for deterministic FAs.}.
Moreover, the implicit $\epsilon$-selfloop $q \xrightarrow{\epsilon} q$ can be viewed as the special
  \emph{0-step-length} transition $q \xrightarrow[0]{\epsilon:\top} q$.
 The \emph{step-length} of $S$ is defined as the maximum of the step-lengths of the symbolic transitions in $S$.
\end{definition}

  \par
   The \emph{symbolic transition} $q \xrightarrow[k]{\sigma:\varphi} \widehat{q}$  represents the set of
   \emph{concrete transitions}
  \begin{equation} \label{eq:<T>}
   \Big \{q \xrightarrow{\sigma\langle a_{1}, a_{2}, \ldots, a_{k}\rangle} \widehat{q} \Big |
   \langle a_{1}, a_{2}, \ldots, a_{k}\rangle \in \llbracket \varphi \rrbracket   \Big \},
  \end{equation}
  where  $\sigma\langle a_{1}, a_{2}, \ldots, a_{k}\rangle$ is the \emph{parameterized event}
  of the concrete transition.
  For example, suppose $X=\{0, 1, 2\}$, then the symbolic transition
  $q \xrightarrow[3]{\sigma:x_{\sigma}^{1} + x_{\sigma}^{2} < x_{\sigma}^{3}} \widehat{q}$
  denotes the set of concrete transitions
  $\{q \xrightarrow{\sigma\langle0,0,1\rangle} \widehat{q}$,
  $q \xrightarrow{\sigma\langle0,0,2\rangle} \widehat{q}$,
  $q \xrightarrow{\sigma\langle0,1,2\rangle} \widehat{q}$,
  $q \xrightarrow{\sigma\langle1,0,2\rangle} \widehat{q}\}$.

\par
According to Definitions \ref{def:efa} and \ref{def:EP-EFA}, EP-EFAs model is just a special type of EFAs model
without state parameters.
This makes it impossible to keep information in the states, and thus limits
 the expressiveness of EP-EFAs model inevitably.

\par
The definitions of the \emph{parameterized string}, \emph{data string},
    \emph{flat data string} in EP-EFAs model are the same as themselves in EFAs model.

\par

\begin{definition} \label{def:longtransition2}
   Given an EP-EFA $S = (Q, \Sigma, X, Q_{0}, Q_{m},  T )$.
   If there exist a series of concrete transitions
  $q_{i} \xrightarrow{v_{i}} q_{i+1}$
  where these  $v_{i}$ are parameterized events,
  $i \in [1:n]$, $n \geq 1$,
  we define the \emph{combined concrete transition} as the path
  $q_{1} \xrightarrow{u} q_{n+1}$ where
   $u = v_{1}v_{2}\ldots v_{n}$.
\end{definition}

 \par
 The language between the set of source states $Q_{1} \subseteq Q$ and
 the set of target states $Q_{2} \subseteq Q$ of the EP-EFA $S$ is defined as follows.
  \begin{align} \label{eq:spf-efalanguage}
  L_{Q_{1}}^{Q_{2}}(S)= \big \{ u | \exists q_{1}, q_{2} \in Q_{1} \text{ s. t. }
  q_{1} \xrightarrow{u} q_{2}\big \}.
  \end{align}
 The \emph{generated language} and \emph{marked language} by the EP-EFA $S$ are, respectively,  defined as
  \begin{equation}
  L(S) = L_{Q_{0}}^{Q}(S) \text{ and } L_{m}(S) = L_{Q_{0}}^{Q_{m}}(S).
    \end{equation}
 The \emph{data language} and \emph{marked data language} of the EP-EFA $S$ are, respectively, defined as
\begin{equation}
L_{d}(S) = \bigcup_{u \in L(S)} \mathfrak{D}(u) \text{ and }
L_{md}(S) = \bigcup_{u \in L_{m}(S)} \mathfrak{D}(u).
\end{equation}
The \emph{flat data language} and \emph{flat marked data language} of the EP-EFA $S$ are, respectively, defined as
\begin{equation}
L_{fd}(S) = \bigcup_{u \in L(S)} \mathfrak{fD}(u) \text{ and }
L_{fmd}(S) = \bigcup_{u \in L_{m}(S)} \mathfrak{fD}(u).
\end{equation}

\begin{definition} \label{def:eq}
 An EP-EFA $S$ and an EFA $E$ are said to be \emph{data-equivalent}, if
  $ L_{fmd}(S) =L_{fmd}(E)$.
\end{definition}

\begin{example} \label{ex:2}
  The EP-EFA $S$ shown in Fig. \ref{fig-2}  also simulates the process of user registering in a web-site.
  In this EP-EFA $S$, the event $\sigma$ carries a 3-tuple of event parameters $\langle x_{\sigma}^{1}, x_{\sigma}^{2}, x_{\sigma}^{3}\rangle$, where the first element
   is for user's nickname, and the second and
  third elements  are both for user's password.
  Hence, if $x_{\sigma}^{2} = x_{\sigma}^{3}$, then the process goes to the final state $q_{1}$ and terminates successfully,
  otherwise it fails and stays in state $q_{0}$.
  It is easy to verify that the EP-EFA $S$ is data-equivalent to the EFA $E$ shown in Fig. \ref{fig-1},
  as the parameters consumed in the transitions $q_{0} \rightarrow q_{1} \rightarrow q_{2} \rightarrow q_{0}$ and
  $q_{0} \rightarrow q_{1} \rightarrow q_{2} \rightarrow q_{3}$ in $E$ are exactly the same as that consumed in
  the transitions $q_{0} \rightarrow q_{0}$ and $q_{0} \rightarrow q_{1}$ in $S$, respectively.
\end{example}
\par
  \begin{figure}[h]
\centering
    \resizebox{0.35\textwidth}{!}{%
             \begin{tikzpicture}[font=\Large,->,>=stealth',shorten >=1pt,auto,node distance=2.5cm,semithick,
                                          every state/.style={font=\large},
                                          new-qs/.style={font=\Large},
                                          new-qs2/.style={font=\Large}]
                        \node[initial, state,new-qs] (q0) {$ q_{0} $};
                        \node[state,new-qs] [right of = q0] (q1) {$ q_{1} $};
                        \node[new-qs2]  [right of = q1] (q5) {$\vdots$};
                        \node[new-qs2] [node distance=0.5cm] [below of = q5] (q6) {$\vdots$};
                        \node[new-qs2] [node distance=0.5cm] [above of = q5] (q7) {$\vdots$};
                       \node[new-qs2] [node distance=0.5cm] [below of = q6] (q8) {$\vdots$};
                       \node[new-qs2] [node distance=0.5cm] [above of = q7] (q9) {$\vdots$};
                        \node[state]  [node distance=3.8cm] [above  of = q5] (q2) {$ q_{2}^{x_{1}} $};
                        \node[state]  [node distance=3.8cm] [below  of = q5] (q4) {$ q_{2}^{x_{K}} $};
                        \node[accepting, state,new-qs]  [right of = q5] (q3) {$q_{3} $};

                    \draw[every node/.style={font=\large}]
                       (q0) [above]  edge  node{$X$} (q1)
                       (q2) [above]  edge  node{$x_{1}$} (q3)
                       (q4) [above]  edge  node{$x_{K}$} (q3)
                       (q1) [below,left]  edge  node{$x_{1}$} (q2)
                       (q1) [below,right]  edge  node{$x_{K}$} (q4)
                       (q4) [below left, bend left]  edge   node{$X \backslash x_{K} $} (q0)
                       (q2) [above left, bend right]  edge  node{$X \backslash x_{1} $} (q0);                            \end{tikzpicture}
             }
\caption{The FA  simulating the registering process in the finite event space $X$ where $x_{i} \in X$, $i \in [1:K]$,
 $K = |X|$, and the states $q_{2}^{x_{j}}$ and the transitions $q_{1} \xrightarrow{x_{j}} q_{2}^{x_{j}}$,
$q_{2}^{x_{j}} \xrightarrow{x_{j}} q_{3}$, and $q_{2}^{x_{j}} \xrightarrow{X \backslash x_{j}} q_{0}$, $j \in [2:K-1]$,
are omitted.}
\label{fig-3}
\end{figure}
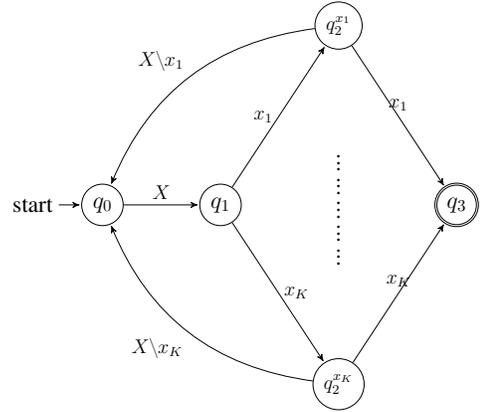

 \begin{remark} \label{remark:1}
 Examples (\ref{ex:1}-\ref{ex:2}) show that,
  although the state parameter is removed,
 EP-EFAs model still retains a fair expressiveness
 by reading multiple event parameters as needed in each transition.
  By the definitions, the models of EFAs and EP-EFAs allow for \emph{infinite} state/event spaces,
  while FAs model only supports \emph{finite} ones.
  This means the parametric models  are more powerful than FAs model.
   In Examples (\ref{ex:1}-\ref{ex:2}),
  suppose $|\Omega| = M$ and $X=\Omega^{\leq N}$, then  $|X| = \sum_{i=1}^{N} M^{i}$.
   To simulate the process of user registering in this finite  space,
   FAs model needs at least $(|X|+3)$ states and $|X|*(|X|+2)$ transitions, as shown in Fig. \ref{fig-3}.
   This suggests that even in a finite space, FAs model may be quite inefficient for certain complex systems
   when compared with the parametric models.
 \end{remark}

\par

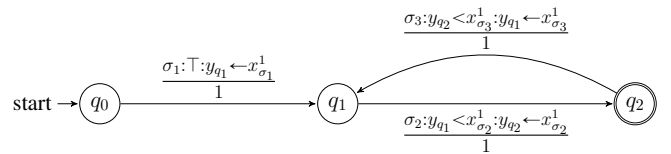
\begin{figure}[h]
\centering
    \resizebox{0.48\textwidth}{!}{%
             \begin{tikzpicture}[font=\Large,->,>=stealth',shorten >=1pt,auto,node distance=6cm,semithick,
                                          every state/.style={font=\Large}]
                \node[initial, state] (q0) {$ q_{0} $};
                \node[state] [node distance=5.2cm] [right of = q0] (q1) {$q_{1} $};
                \node[accepting, state] [node distance=6.5cm] [right of = q1] (q2) {$q_{2} $};

            \draw[every node/.style={font=\LARGE}]
                   (q0) [above]  edge  node{$\frac{\sigma_{1}:\top:y_{q_{1}}\leftarrow x_{\sigma_{1}}^{1}}{1}$} (q1)
                   (q1) [below]  edge  node{$\frac{\sigma_{2}:y_{q_{1}}<x_{\sigma_{2}}^{1}:y_{q_{2}}\leftarrow x_{\sigma_{2}}^{1}}{1}$} (q2)
                   (q2) [above,bend right]  edge  node{$\frac{\sigma_{3}: y_{q_{2}}<x_{\sigma_{3}}^{1}:
                           y_{q_{1}} \leftarrow x_{\sigma_{3}}^{1}}{1}$} (q1);
               \end{tikzpicture}
        }
\caption{The EFA $E$ with $X=Y=\mathbb{N}$  accepting even number of increasing natural numbers. }
\label{fig-4}
\end{figure}

\begin{proposition} \label{pro:1}
 There exists an EFA $E$ that cannot be data-equivalent with any EP-EFA $S_{E}$.
\end{proposition}
\begin{IEEEproof}
First of all, we construct an EFA $E$ with $X=Y=\mathbb{N}$, as shown in Fig. \ref{fig-4}.
   Obviously, $E$ accepts even number of increasing natural numbers, i.e.,
   the marked data string of $E$ has the form of
  $a_{1}a_{2}\ldots a_{2*n}$, where $n \geq 1$, $a_{i+1} > a_{i}$, $i \in [1:(2*n-1)]$.
  \par

  Secondly,
   we prove there does not exist a data-equivalent EP-EFA $S_{E}$ for  the EFA $E$ by contradiction.
  Suppose there exists a data-equivalent EP-EFA $S_{E}$, where the number of the states is $m$ and
  the step-length is $K$.
  Take a flat marked data string of $S_{E}$ $u=a_{1}a_{2}\ldots a_{2*n}$ where $2*n > (m-1)*K$.
  Suppose that $u$ visits the sequence of
  states $q_{0}\rightarrow q_{1} \rightarrow \ldots \rightarrow q_{l}$  in $S_{E}$,
  where $q_{0} \in Q_{0}$ and $q_{l} \in Q_{m}$.
  Since the step-length of $S_{E}$ is $K$, we have $l*K \geq 2*n$, and thus $l>m-1$.
This means  that there exist  two states $q_{i}, q_{j}$ in the sequence of visited states of $u$
 such that $q_{i} = q_{j}$ and $0 \leq i < j \leq l$,  as the EP-EFA $S_{E}$ has $m$ states.
 Suppose that the parameters consumed from state $q_{i}$ to state $q_{j}$ are $a_{\widehat{i}}a_{\widehat{i}+1}\ldots a_{\widehat{j}}$,
  $1  \leq \widehat{i} < \widehat{j} \leq 2*n$.
  Obviously, the flat data string $\widehat{u} = a_{1}\ldots a_{\widehat{i}-1} a_{\widehat{i}}a_{\widehat{i}+1}\ldots a_{\widehat{j}} a_{\widehat{i}}a_{\widehat{i}+1}\ldots a_{\widehat{j}} a_{\widehat{j}+1}\ldots a_{2*n}$
   also can be marked by $S_{E}$.
  Since $S_{E}$ and $E$ are data-equivalent, $\widehat{u}$ is marked by $E$.
  However, it is not true, as $a_{\widehat{j}} > a_{\widehat{i}}$ and
   $\hat{u}$ is not a sequence of increasing numbers.
   Hence, the contradiction is generated, which implies there does \emph{not} exist a data-equivalent EP-EFA $S_{E}$ for the EFA $E$ shown in Fig. \ref{fig-4}.
      \end{IEEEproof}

\begin{proposition} \label{pro:2}
  For any EP-EFA $S$, there always exists a data-equivalent EFA $E_{S}$.
\end{proposition}
\begin{IEEEproof}
It is straightforward by Definitions \ref{def:efa} and \ref{def:EP-EFA}.
   \end{IEEEproof}

\begin{remark}
     Propositions \ref{pro:1} and \ref{pro:2} imply that EFAs model is more expressive than EP-EFAs model.
     The models of EFAs and EP-EFAs extend FAs to an infinite model
     by means of the symbolic transitions carrying predicates over the infinite parameter space.
     With the help of the satisfiability modulo theories (SMT) solvers (e.g., Z3,  Open SMT, MathSAT5, etc., see  \cite{solver2} for details),
     the data types  that can be efficiently processed by parametric  models include
     real/integer, bit vectors, arrays, difference logic, inductive data, etc.
      Therefore, the models of EFAs and EP-EFAs are quite expressive tools for DESs.
\end{remark}

\par
A longer step-length adds the expressiveness of EP-EFAs.
As evidence, the $k$-step-length transition
$q_{1} $ $\xrightarrow[k]{\sigma: x_{\sigma}^{1} < x_{\sigma}^{2} < \ldots < x_{\sigma}^{k}} q_{2}$  has no equivalent series of transitions with a lower step-length.
However, for the EFAs model, a longer step-length does \emph{not} add its expressiveness,
as the state parameter can be used to store the necessary information during the transitions.
The subsequent proposition presents a formal demonstration for this fact.
\begin{proposition} \label{pro:3}
  For any $m$-step-length EFA $E_{m}$, $m > 1$, there always exists a data-equivalent 1-step-length EFA $E_{1}$.
\end{proposition}
\begin{IEEEproof}
  Given any $m$-step-length EFA $E_{m}$, we construct the data-equivalent $1$-step-length EFA $E_{1}$ as follows.
  For each symbolic transition $(q,y_{q} )\xrightarrow[k]{\sigma:\varphi:\xi} (\widehat{q}, y_{\widehat{q}})$
  of $E_{m}$, $1 < k \leq m$,
  we add $(k-1)$ new states: $q^{i}$, $i\in [1:k-1]$, and $k$ events $\sigma^{j}$, $j \in [1:k]$,
  and then construct a chain of $k$ 1-step-length transitions $q^{j-1} \xrightarrow[1]{\sigma^{j}:\varphi^{j}:\xi^{j}} q^{j}$ where $q^{0} = q$ and $q^{k} = \widehat{q}$
  to replace the transition $q \xrightarrow[k]{\sigma:\varphi:\xi} \widehat{q}$.
  Specifically, the update functions are defined as follows:
    $\xi^{1} \overset{\text{def}}{=} [ y_{q^{1}}(1) \leftarrow x_{\sigma^{1}}^{1} ] $
    and for $j \in [2:k-1]$,
    $\xi^{j} \overset{\text{def}}{=}
    [ y_{q^{j}}(1) \leftarrow y_{q^{j-1}}(1);
    \ldots;
    y_{q^{j}}(j-1) \leftarrow y_{q^{j-1}}(j-1);
    y_{q^{j}}(j) \leftarrow x_{\sigma^{j}}^{1}
    ]$ where $y_{q^{j}}(i)$ means the $i^{th}$ element of the state parameter of $q^{j}$,
    and $\xi^{k} \overset{\text{def}}{=} \xi(x_{\sigma}^{1}/y_{\widehat{q}}(1), \ldots, x_{\sigma}^{k-1}/y_{\widehat{q}}(k-1), x_{\sigma}^{k}/x_{\sigma^{k}}^{1})$,
    where the ``A/B" denotes the substituting $A$ by $B$ in function $\xi$.
   The predicates are as follows:
   $\varphi^{i} = \top$ for $i\in [1:k-1]$, and
  $\varphi^{k} = \varphi(x_{\sigma}^{1}/y_{\widehat{q}}(1), \ldots, x_{\sigma}^{k-1}/y_{\widehat{q}}(k-1), x_{\sigma}^{k}/x_{\sigma^{k}}^{1})$ where the ``A/B" denotes the substituting $A$ by $B$ in the predicate.
  Obviously, $\varphi^{k}$ is a $(Y \times X)$-predicate where $Y = X^{k-1}$.
  The intuitive meaning of these new transitions is as follows.
  Each new transition is responsible for transmitting state parameters from source state to target state
  and storing one event parameter to target state parameter;
   and the first $(k-1)$ transitions are guarded with $\top$ and
    the last one is guarded with $\varphi^{k}$ that is equivalent with $\varphi$.
    In addition, $\xi^{k}$ is also equivalent with $\xi$.
     This means that for any $k$ event parameters, the transition
     $(q,y_{q} )\xrightarrow[k]{\sigma:\varphi:\xi} (\widehat{q}, y_{\widehat{q}})$
     is fired if and only if the  chain of transitions is fired,
     and meanwhile the parameter of the final state $\widehat{q}$ is also updated in the same way.
   Thus, by replacing each transition of $E_{m}$ with such a chain of transitions,
   we can obtain the data-equivalent 1-step-length $E_{1}$.
\end{IEEEproof}

\section{Problem Formulation and Assumptions}
In this section, we present some assumptions and then formulate the problems discussed in this paper.
\par
In rest of this paper, we focus on the problem of opacity analysis for a parametric DES modeled by
 an EFA 
 $E = (Q, \Sigma, X, Q_{0}, Q_{m}, Y, Y_{0}, R )$
or an EP-EFA 
$S = (Q, \Sigma, X, $ $ Q_{0}, Q_{m}, T )$.  In the following, the parametric DES is denoted by $G$, and
 the notation $L_{Q_{1}}^{Q_{2}}(G)$ is the language calculated by Equation (\ref{eq:efalanguage}) when $G$ is an EFA, and by Equation (\ref{eq:spf-efalanguage}) when $G$ is an EP-EFA.
\par
The basic assumptions in this paper are as follows.
\begin{itemize}

 \item
  \textbf{\emph{Assumption 1}}: The secret and non-secret behavior of the parametric  system can be coded into
   its state space.
    We consider the following two cases:
   1) the secret and non-secret behavior are the sets of data strings arriving in the given secret states $Q_{s}\subseteq Q$ and non-secret states $Q_{ns}\subseteq Q$, respectively, and Definitions \ref{def:cso} and \ref{def:infinite} are of this case;
   2) the secret and non-secret behavior are the sets of data strings originating from the given secret initial states
   $Q_{s} \subseteq Q_{0}$ and non-secret initial states $Q_{ns} \subseteq Q_{0}$, respectively,
   and Definition \ref{def:iso} is of this case.

    \item
  \textbf{\emph{Assumption 2}}: The intruder knows the complete structure of the parametric DES $G$,
  and he can observe the data exchanges between the system and its environment during the interactions
  (i.e., \emph{data language} $L_{d}(G)$) through a \emph{static} observation function $\theta$.
  The observation function $\theta$ is defined as:
  for any data string $ d =  \langle a_{1}^{1}a_{1}^{2}\ldots a_{1}^{k_{1}}\rangle$
  $\langle a_{2}^{1}a_{2}^{2}\ldots a_{2}^{k_{2}}\rangle \ldots \langle a_{j}^{1}a_{j}^{2}\ldots a_{j}^{k_{j}}\rangle \in L_{d}(G)$,
  \begin{align}
    \theta(d) = & \langle\theta(a_{1}^{1})\theta(a_{1}^{2}) \ldots \theta(a_{1}^{k_{1}}) \rangle
 \langle\theta(a_{2}^{1})\theta(a_{2}^{2}) \ldots \theta(a_{2}^{k_{2}}) \rangle \nonumber \\
  &  \ldots \langle \theta(a_{j}^{1})\theta(a_{j}^{2}) \ldots \theta(a_{j}^{k_{j}}) \rangle \nonumber \\
  \text{where } & \theta(a_{m}^{n}) =
 \begin{cases}
            a_{m}^{n},                 & \text{if }  \vartheta(a_{m}^{n}) \text{ holds},\\
           \epsilon,                 & \text{otherwise},\\
          \end{cases}
  \end{align}
   and  $\vartheta$ is the $X$-predicate describing the observable condition for  data elements, and
 the \emph{empty observation} ``$\langle \epsilon \rangle$" in $\theta(d)$ can be removed directly.
 The set of observations for $G$ is defined as
 $\Theta(G) = \bigcup_{u \in L(G)}\theta(\mathfrak{D}(u))$.
\end{itemize}

 \begin{definition}
  An \emph{observable unit} of the observation $w$, $w \in \Theta(G)$, is  the substring of form
   ``$\langle a_{i}a_{i+1}\ldots a_{i+k}\rangle$", $k \geq 0$, in $w$.
  Let $|w|_{u}$ denote the number of observable units in $w$.
 \end{definition}
 \par

 \par

\begin{remark}
  According to the definition,  the observations such as
  ``$\langle a_{1}a_{2}\rangle\langle a_{3} \rangle$" and
  ``$\langle a_{1}\rangle\langle a_{2}a_{3} \rangle$" are considered to be different.
  Two identical observations  have the same number of observable units and the corresponding units are equal to each other.
  Therefore, an observable unit is regarded as a \emph{minimal information structure} acquired by the intruder,
  and this paper considers the data language rather than the flat data language in  opacity analysis.
  The main reasons for this treatment are as follows.
1) Since each parameter tuple is transmitted between the system and its environment as a whole and the observable unit
   is the observable part of parameter tuple, the intruder will obtain each observable unit as a whole.
2) Similar to the literature of opacity analysis \cite{opacity-review}-\cite{output-enforcement3},
this paper also assumes that intruders have sufficient memory and computation capabilities to
 keep the history of the observations and
update the state estimation for the system instantaneously by their latest observations.
\end{remark}
\par

Based on these assumptions, we present three opacity properties for parametric DESs in the following.

\par

\begin{definition} \label{def:cso} (current-state opacity)
  Given the parametric DES $G$ with the set of secret states $Q_{s} \subseteq Q$,
  the set of non-secret states $Q_{ns} \subseteq Q$, and the observation function $\theta$.
   $G$ is said to be \emph{current-state opaque}  w.r.t. $Q_{s}$, $Q_{ns}$ and $\theta$, if
   \begin{align} \label{eq:cso}
     (\forall u \in L^{Q_{s}}_{Q_{0}}(G)) (\exists v \in L^{Q_{ns}}_{Q_{0}}(G))
      \theta(\mathfrak{D}(u))=\theta(\mathfrak{D}(v)).
   \end{align}
\end{definition}

\par
\begin{definition} \label{def:iso} (initial-state opacity)
  Given the parametric DES $G$ with the set of secret initial states $Q_{s} \subseteq Q_{0}$,
  the set of non-secret initial states $Q_{ns} \subseteq Q_{0}$,
  and the observation function $\theta$.
   $G$ is said to be \emph{initial-state opaque}  w.r.t. $Q_{s}$, $Q_{ns}$ and $\theta$, if
   \begin{align} \label{eq:iso}
     (\forall u \in L^{Q}_{Q_{s}}(G)) (\exists v \in L^{Q}_{Q_{ns}}(G))
     \theta(\mathfrak{D}(u))=\theta(\mathfrak{D}(v)).
   \end{align}
\end{definition}

\par

\begin{definition} \label{def:infinite} (infinite-step opacity)
  Given the parametric DES $G$ with the set of secret states $Q_{s} \subseteq Q$,
  the set of non-secret states $Q_{ns} \subseteq Q$,
  and  observation function $\theta$.
   $G$ is said to be \emph{infinite-step opaque}  w.r.t. $Q_{s}$, $Q_{ns}$ and $\theta$, if
 \begin{align} \label{eq:infinite}
&(\forall u\widehat{u} \in L_{Q_{0}}^{Q}(G): u \in L^{Q_{s}}_{Q_{0}}(G))
(\exists v\widehat{v} \in L_{Q_{0}}^{Q}(G) : v \in L^{Q_{ns}}_{Q_{0}}(G))\nonumber \\
 &[\theta(\mathfrak{D}(u))=\theta(\mathfrak{D}(v)) \wedge
                       \theta(\mathfrak{D}(\widehat{u}))=\theta(\mathfrak{D}(\widehat{v}))].
 \end{align}
\end{definition}

\par

The opacity properties of parametric DESs presented in Definitions \ref{def:cso}, \ref{def:iso}, \ref{def:infinite}
have the same intuitive meanings as their counterparts of the  classic DESs.
We would  investigate the opacity properties for  EFAs and EP-EFAs in Sections IV and V, respectively.

\section{Undecidability of Opacity in EFAs}
In this section, we prove that the opacity properties presented in Definitions \ref{def:cso}, \ref{def:iso}, \ref{def:infinite} for  EFAs are all undecidable in general.
The main idea of the proof is reducing
the halting problem of two-counter machines to the verification of the opacity properties.

\par

A counter machine is an abstract machine used to model computation in formal logic and theoretical computer science.
A counter machine consists of several registers, each of which only can store an integer number,
and a set of arithmetic operations and control instructions.
Minsky introduced a type of counter machines including two registers $r_{j}$, $j \in \{1, 2\}$, and three instructions: $INC(r_{j})$, $DEC(r_{j})$ and $JZ(r_{j}, z)$ with the semantics of $r_{j} \leftarrow r_{j} + 1$, $r_{j} \leftarrow r_{j} - 1 $, and
$goto(z)$ if $r_{j}=0$, respectively \cite{CM}.
This kind of machines is usually called \emph{Two-Counter Machines (2CMs)} in the literature.

\par
\begin{lemma} \label{lemma-1}
2CMs are Turing equivalent \cite{CM}.
\end{lemma}
\par
 It is well known that the halting problem for Turing machines is undecidable.
 Therefore, by Lemma 1, we have the following result.
\begin{lemma} \label{lemma-2}
  The halting problem of 2CMs is undecidable.
\end{lemma}

\par
Obviously, a configuration of a 2CM with program $P$ can be described as a triple
 $(r_{1},r_{2}, c ) \in \mathbb{N}^{3}$,
where $r_{1}$ and $r_{2}$ keep the values of the first and second registers, respectively,
and $c$ keeps the value of program counter.
Let $\overline{x}(j)$ denote the $j^{th}$ entry of the configuration $\overline{x} \in \mathbb{N}^{3}$,  $j\in [1:3]$.
Let $|P|$ denote the number of instructions in program $P$.

\par
Firstly, we formulate the ($\mathbb{N}^{3} \times \mathbb{N}^{3}$)-predicate $\varphi^{step}$ that characterizes the configuration evolution of the 2CM with program $P$ after executing a single instruction, where the first and second
elements refer to the current and subsequent configurations, respectively.
Let $\varphi_{i}$ be the ($\mathbb{N}^{3} \times \mathbb{N}^{3}$)-predicate describing the relation
of the configurations before and after the executing of the $i^{th}$ instruction of program $P$.
We formulate $\varphi_{i}$ according to the type of the $i^{th}$ instruction as follows.
\begin{enumerate}
  \item
  If the $i^{th}$ instruction is $INC(r_{j})$, $j \in \{1,2\}$, then
\begin{align*}
 \varphi_{i}(\overline{y}, \overline{x}) \overset{\text{def}}{=} [(\overline{x}(j) = \overline{y}(j) + 1) \wedge   (\overline{x}(3-j) =   \nonumber \\
 \overline{y}(3-j)) \wedge (\overline{y}(3) = i ) \wedge (\overline{x}(3) = i + 1)],
\end{align*}
where the first clause means that the $j^{th}$ register increases by 1,  the second clause means the other register
remains unchanged, the third and fourth clauses mean that the program is executing the $i^{th}$ instruction and the next  instruction to be executed is the $(i+1)^{th}$ one, respectively.

  \item
   If the $i^{th}$ instruction is $DEC(r_{j})$, $j \in \{1,2\}$, then
\begin{align*}
 \varphi_{i}(\overline{y}, \overline{x}) \overset{\text{def}}{=} [(\overline{x}(j) = \overline{y}(j) - 1) \wedge   (\overline{x}(3-j) =  \nonumber \\
  \overline{y}(3-j)) \wedge (\overline{y}(3) = i ) \wedge (\overline{x}(3) = i + 1)].
\end{align*}
The intuitive meaning of this equation is  similar to that of the previous one.

\item
If the $i^{th}$ instruction is $JZ(r_{j},z)$, $j \in \{1,2\}$, then
\begin{align*}
\varphi_{i}(\overline{y}, \overline{x}) \overset{\text{def}}{=} [(\overline{x}(1) = \overline{y}(1) ) \wedge   (\overline{x}(2) = \overline{y}(2)) \wedge  \nonumber \\
(\overline{y}(3) = i ) \wedge  (\overline{x}(3) = ( \overline{y}(j) = 0 \quad ? \quad z: i+1 ))],
\end{align*}
where the first and second causes mean that both the registers remain unchanged,
the third cause means that the program is executing the $i^{th}$ instruction, and
the last cause adopts a Java-language-style expression to describe the if-then-else term,
i.e., if the register $r_{j}$ equals 0, then the next instruction to be executed is the $z^{th}$ one,
otherwise the $(i+1)^{th}$ one.
\end{enumerate}
Hence, we obtain the special predicate $\varphi^{step}$ for program $P$ as follows.
\begin{equation} \label{equ:step}
  \varphi^{step}(\overline{y}, \overline{x}) \overset{\text{def}}{=} \bigvee_{i \in [1:|P|]} \varphi_{i}(\overline{y}, \overline{x}).
\end{equation}
 The ($\mathbb{N}^{3} \times \mathbb{N}^{3}$)-predicate $\varphi^{eq}$ describing whether
two configurations are equal to each other or not is defined as follows.
\begin{align} \label{equ:eq}
  \varphi^{eq}(\overline{y}, \overline{x}) \overset{\text{def}}{=}
  \bigwedge_{i\in\{1,2,3\}}[(\overline{x}(i) = \overline{y}(i) )]
\end{align}
Obviously, $\varphi^{step}$ and $\varphi^{eq}$ are both predicates
in the Boolean algebra
$\mathcal{A}=(\mathbb{N}^{3} \times \mathbb{N}^{3}, \Psi, \llbracket \bullet \rrbracket)$.
For the specific program $P$,
we denote by $\mathbb{N}^{3}$-predicates $\varphi^{ini}$ and $\varphi^{fin}$ its initial configuration
and final configuration, respectively.

\par
Based on the above discussions, we prove that the current-state opacity of  EFAs is
undecidable by constructing a special parametric DES $E_{P}$ w.r.t program $P$ and reducing the halting problem
 of $P$ to the verification of current-state opacity of $E_{P}$.
\par

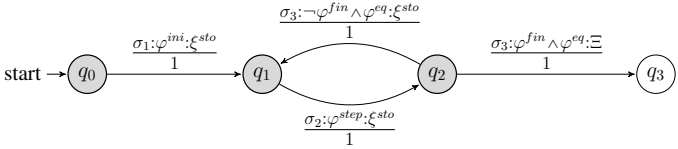
\begin{figure}[h]
\centering
    \resizebox{0.5\textwidth}{!}{%
     \begin{tikzpicture}[font=\Large,->,>=stealth',shorten >=1pt,auto,node distance=6.5cm,semithick,
                                  every state/.style={font=\Large},
                                  new-qs/.style={fill=gray!30,thick,font=\Large}]
                \node[initial, state, new-qs] (q0) {$ q_{0} $};
                \node[state,new-qs] [node distance=4cm] [right of = q0] (q1) {$q_{1} $};
                \node[state,new-qs] [node distance=4cm] [right of = q1] (q2) {$q_{2} $};
                \node[state,state] [node distance=5.0cm] [right of = q2] (q3) {$q_{3} $};

            \draw[every node/.style={font=\LARGE}]
             (q0) [above]  edge  node{$\frac{\sigma_{1}:\varphi^{ini}:\xi^{sto}}{1}$} (q1)
              (q2) [above] edge  node{$\frac{\sigma_{3}: \varphi^{fin} \wedge \varphi^{eq}:\Xi}{1}$}  (q3)
             (q1) [below, bend right] edge  node{$\frac{\sigma_{2}: \varphi^{step} :\xi^{sto}}{1}$} (q2)
             (q2) [above, bend right] edge  node{$\frac{\sigma_{3}: \neg \varphi^{fin} \wedge \varphi^{eq}:\xi^{sto}}{1}$ } (q1);
               \end{tikzpicture}
            }
\caption{The EFA $E_{P}$ for a 2CM with program $P$, where the domains of  event parameter $X$ and  state parameter
  $Y$ are both $\mathbb{N}^{3}$.}
 \label{fig-5}
\end{figure}

\begin{theorem} \label{thm:cso}
  The current-state opacity of  EFAs is undecidable in general.
\end{theorem}
\begin{IEEEproof}
Firstly,  we construct the EFA
$E_{P} = \big \{ Q=\{ q_{0},q_{1}, $ $ q_{2},q_{3}\}, \Sigma = \{\sigma_{1}, \sigma_{2}, \sigma_{3}, \sigma_{4}\}, X = \mathbb{N}^{3}, Q_{0}=\{q_{0}\}, Y = \mathbb{N}^{3}, Y_{0}=\llbracket \varphi^{ini} \rrbracket , R \big \}$ w.r.t. a 2CM with program $P$
(shown in Fig. \ref{fig-5}).
The predicates of $\varphi^{step}$ and $\varphi^{eq}$ are  defined in
Equations (\ref{equ:step}) and (\ref{equ:eq}), respectively.
The predicates of $\varphi^{ini}$ and $\varphi^{fin}$, as the logic characterization for the
initial and final configurations of program $P$, respectively, are $X$-predicates and also can be regarded as special
$(Y\times X)$-predicates where the first variable (i.e., state parameter) has no influence to the predicates.
In the symbolic transitions, the update function $\xi^{sto}$ just stores the event parameter to
the target state as parameter,
e.g., $\xi^{sto}$ in the transition from $q_{0}$ to $q_{1}$ is defined as: $y_{q_{1}} \leftarrow x_{\sigma_{1}}^{1}$.
Let the set of secret states be $Q_{s} = \{ q_{0},q_{1},q_{2} \}$ and
the set of non-secret states be $Q_{ns} = \{ q_{3} \}$.
Consider the observation function $\theta$: $\forall u \in (\mathbb{N}^{3})^{*}$, $\theta(u) = \epsilon$.
According to Definition \ref{def:cso}, the parametric DES $E_{P}$  is current-state opaque if and
only if the non-secret behavior is non-empty, i.e., the state $q_{3}$ is reachable from the initial state $q_{0}$.

\par

According to Fig. \ref{fig-5}, the data strings (i.e., the sequence of configurations) that can reach the state $q_{3}$
from the initial state $q_{0}$ have the form of
$v = a_{1}a_{2}\ldots a_{2*n}a_{2*n+1}$, $n \geq 1$,
and $q_{3}$ is reachable if and only if  $v$
satisfies the following formulae:
$a_{1} \in \llbracket \varphi^{ini} \rrbracket$, $a_{2*n+1} \in \llbracket \varphi^{fin} \rrbracket$,
$a_{2*j+1} \not\in \llbracket \varphi^{fin} \rrbracket$, $j \in [1:n-1]$,
and for $i \in [1:n]$,
$(a_{2*i-1},a_{2*i}) \in \llbracket \varphi^{step} \rrbracket $ and
$(a_{2*i},a_{2*i+1}) \in \llbracket \varphi^{eq} \rrbracket $.

\par
For such sequence $v$ satisfying aforementioned formulae, there exists a one-to-one corresponding sequence
$w=a_{1}a_{2}a_{4} $ $ \ldots a_{2*(n-1)}a_{2*n}$, $n \geq 1$,  where $a_{1}$ and $a_{2*n}$ are, respectively,
the initial and final configurations, and each pair of adjacent configurations satisfies the predicate $\varphi^{step}$.
This means that $w$ is exactly the evolution sequence of configurations during the execution of program $P$,
i.e., the 2CM with program $P$ halts if and only if there exists such sequence $w$.

\par
By Lemma \ref{lemma-2},  the halting problem of 2CMs is undecidable,
which implies the undecidability of the existence of such $w$,
and further implies  the undecidability of the existence of such $v$.
Hence, the reachability of state $q_{3}$ in $E_{P}$ is undecidable,
and so is the current-state opacity of $E_{P}$.
Therefore, the current-state opacity of  EFAs is undecidable in general.
   \end{IEEEproof}

\par

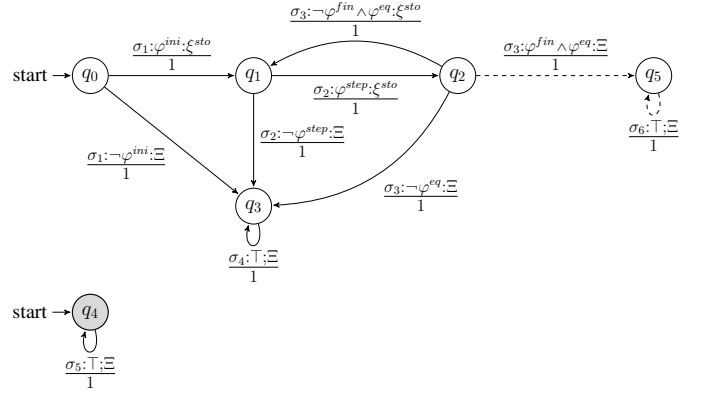
\begin{figure}[h]
\centering
    \resizebox{0.5\textwidth}{!}{%
     \begin{tikzpicture}[font=\Large,->,>=stealth',shorten >=1pt,auto,node distance=7.5cm,semithick,
                                  every state/.style={font=\Large},
                                  new-qs/.style={fill=gray!30,thick,font=\Large}]
                \node[initial, state] (q0) {$ q_{0} $};
                \node[state] [node distance=4cm] [ right of = q0] (q1) {$q_{1} $};
                \node[state] [node distance=5cm] [right of = q1] (q2) {$q_{2} $};
                \node[state] [node distance=4.8cm] [right of = q2] (q5) {$q_{5} $};
                \node[state] [node distance=3.2cm] [below  of = q1] (q3) {$q_{3} $};
                \node[initial, state, new-qs] [node distance=5.8cm] [below of = q0] (q4) {$q_{4} $};

            \draw[every node/.style={font=\LARGE}, eg/.style={dashed}]
             (q0) [above]  edge  node{$\frac{\sigma_{1}:\varphi^{ini}:\xi^{sto}}{1}$} (q1)
             (q0) [below left]  edge  node{$\frac{\sigma_{1}:\neg \varphi^{ini}:\Xi}{1}$} (q3)
             (q1)  [below] edge  node{$\frac{\sigma_{2}: \varphi^{step} :\xi^{sto}}{1}$} (q2)
             (q1) [right,right] edge  node{$\frac{\sigma_{2}: \neg \varphi^{step} :\Xi}{1}$} (q3)
             (q2) [above ] edge[eg]  node{$\frac{\sigma_{3}: \varphi^{fin} \wedge \varphi^{eq}:\Xi}{1}$}  (q5)
             (q5) edge[loop below,eg] node{$\frac{\sigma_{6}:\top;\Xi}{1}$} ( )
             (q3) edge[loop below] node{$\frac{\sigma_{4}:\top;\Xi}{1}$} ( )
             (q4) edge[loop below] node{$\frac{\sigma_{5}:\top;\Xi}{1}$} ( )
             (q2) [below right, bend left] edge  node{$\frac{\sigma_{3}: \neg \varphi^{eq}:\Xi}{1}$ } (q3)
             (q2) [above, bend right] edge  node{$\frac{\sigma_{3}: \neg \varphi^{fin} \wedge \varphi^{eq}:\xi^{sto}}{1}$ } (q1);
               \end{tikzpicture}
            }
\caption{The EFAs $\widehat{E}_{P}$ and $\widetilde{E}_{P}$ for a 2CM with program $P$,
where the sets of states in $\widehat{E}_{P}$ and $\widetilde{E}_{P}$ are $\{q_{0},q_{1},q_{2},q_{3},q_{4}\}$
and $\{q_{0},q_{1},q_{2},q_{3},q_{5}\}$, respectively.
}
\label{fig-6}
\end{figure}
\par
\begin{theorem} \label{thm:iso}
  The initial-state opacity of  EFAs is undecidable in general.
\end{theorem}
\begin{IEEEproof}
First of all, we construct the EFA
$\widehat{E}_{P} = \big \{ Q=\{ q_{0}, \ldots,q_{4}\},  \Sigma = \{\sigma_{1}, \ldots ,\sigma_{5}\}, X=\mathbb{N}^{3}, Q_{0}=\{q_{0},q_{4}\}, Y=\mathbb{N}^{3}, Y_{0}=\llbracket \varphi^{ini} \rrbracket , R \big \}$ for a 2CM with program $P$.
In EFA $\widehat{E}_{P}$, the predicates $\varphi^{ini}$, $\varphi^{eq}$, $\varphi^{step}$ and $\varphi^{fin}$,
and update function $\xi^{sto}$ have the same definitions as themselves in $E_{P}$ (shown in Fig. \ref{fig-5}).
Let the set of secret initial states be $Q_{s} = \{ q_{4} \}$ and
the set of non-secret initial states be $Q_{ns} = \{ q_{0} \}$.
Consider the observation function $\theta$: $\theta(u) = u$, $ u \in (\mathbb{N}^{3})^{*}$.
Under these settings, we have the following fact.
\begin{equation} \label{eq:secret-observation}
    \bigcup\nolimits_{v \in L_{Q_{s}}^{Q}(\widehat{E}_{P})} \theta(\mathfrak{D}(v)) =
    \bigcup\nolimits_{v \in L_{\{q_{4}\}}^{\{q_{4}\}}(\widehat{E}_{P})} \theta(\mathfrak{D}(v)) = ({\mathbb{N}^{3}})^{*}.
\end{equation}
That is, the set of  observations for \emph{secret} behavior is the universal set $({\mathbb{N}^{3}})^{*}$.
According to Definition \ref{def:infinite}, $\widehat{E}_{P}$ is initial-state opaque if and only if
the set of the observations for \emph{non-secret} behavior is also the universal set $({\mathbb{N}^{3}})^{*}$, i.e.,
\begin{multline}  \label{eq:nonsecret-observation}
\bigcup\nolimits_{u \in L_{Q_{ns}}^{Q}(\widehat{E}_{P})} \theta(\mathfrak{D}(u)) =   \\
    \bigcup\nolimits_{u \in L_{\{q_{0}\}}^{\{q_{0},q_{1},q_{2},q_{3}\}}(\widehat{E}_{P})}\theta(\mathfrak{D}(u))  = ({\mathbb{N}^{3}})^{*}.
\end{multline}

\par
In order to investigate the validness of Equation (\ref{eq:nonsecret-observation}),
we construct a new EFA $\widetilde{E}_{P}$ from $E_{\widehat{P}}$ by removing state $q_{4}$ and its corresponding transitions,
and adding a state $q_{5}$ and two corresponding transitions
(i.e., the transitions denoted by dotted-arrow in Fig. \ref{fig-6}).
 In the new EFA $\widetilde{E}_{P}$, we have the fact that
 the disjunction of the predicates in the transitions originating the same state is equal to the true predicate $\top$, e.g., for state $q_{2}$,
 $ (\neg \varphi^{fin} \wedge \varphi^{eq} ) \vee (\varphi^{fin} \wedge \varphi^{eq}) \vee ( \neg \varphi^{eq}) = \top$.
Hence, we have the fact that
\begin{equation} \label{eq:universe}
\bigcup\nolimits_{u \in L_{\{q_{0}\}}^{\{q_{0},q_{1},q_{2},q_{3},q_{5}\}}(\widetilde{E}_{P})}\theta(\mathfrak{D}(u))  = ({\mathbb{N}^{3}})^{*}.
\end{equation}
According to Equation (\ref{eq:universe}),
it is obvious that Equation (\ref{eq:nonsecret-observation}) holds if and only if the state $q_{5}$ is \emph{not} reachable in $\widetilde{E}_{P}$.
Notice that the reachability of  $q_{5}$ in $\widetilde{E}_{P}$ is identical to
the reachability of $q_{3}$ in $E_{P}$ (shown in Fig. \ref{fig-5}), which has been proved to be undecidable in Theorem \ref{thm:cso}.
Hence, the validness of Equation (\ref{eq:nonsecret-observation}) is undecidable,
and so is the initial-state opacity of EFA $\widehat{E}_{P}$.
Therefore, the initial-state opacity of  EFAs is undecidable in general.
   \end{IEEEproof}

\par

\begin{theorem} \label{thm:infinite}
  The infinite-step opacity of  EFAs is undecidable in general.
\end{theorem}
\begin{IEEEproof}
  We consider the same EFA $E_{P}$ with the same secret states, non-secret states and the observation function
  as that in Theorem \ref{thm:cso}.
  By Definition \ref{def:infinite}, $E_{P}$ is infinite-step opaque if and only if the state $q_{3}$
  is reachable from the initial state $q_{0}$, which has been proved to be undecidable in Theorem \ref{thm:cso}.
  Therefore, infinite-step opacity of $E_{P}$ is undecidable, and
  infinite-step opacity of  EFAs is undecidable in general.
   \end{IEEEproof}

   \par
\begin{remark}
As mentioned before,
EFAs model is a quite powerful tool to simulate the interactions between a system and its environment.
However, the coexistence of event and state parameters in the predicates complicates this model and
make the properties of opacity undecidable.
Hence, it is necessary to consider the EP-EFAs model where the state parameter is removed.
\end{remark}

\section{Opacity of  EP-EFAs}

In this section, we investigate the current-state opacity, initial-state opacity and infinite-step opacity
of  EP-EFAs.
We present the verification algorithms for these opacity properties firstly,
and then analyze the complexity of these algorithms.

\subsection{Current-State Opacity of EP-EFAs}
In fact, Definition \ref{def:cso} implies that
 the current-state opacity holds if and only if for any observation,
 the intruder cannot determine the system is in the secret states.
 For the convenience of demonstrating this issue, we present the following notion.
 \begin{definition} \label{def:est}
 Given the EP-EFA $S=  (Q, \Sigma, X,  Q_{0}, T)$,
 the state estimation function $Est^{S}: \Theta(S) \rightarrow 2^{Q}$  is defined as follows:
 for any observation $ w \in \Theta(S)$,
 \begin{multline}\label{eq:est}
   Est^{S}(w) = \{ q \in Q | \exists q_{0} \in Q_{0}, u \in L(S), \text{ s.t. } q_{0} \xrightarrow{u} q   \\
    \wedge w = \theta(\mathfrak{D}(u)) \}.
 \end{multline}
 \end{definition}

\par
For classic DESs, the state estimations  can be calculated by constructing a special  automaton: \emph{observer} \cite{cso}.
 Inspired by this idea, we present an algorithm (Algorithm \ref{alg:1}) to construct the \emph{symbolic observer}
 $Obs(S) = \{ Q^{obs}, q^{obs}_{0},  T^{obs}\}$ for the EP-EFA $S=  (Q, \Sigma, X, Q_{0}, T)$.
 The symbolic observer $Obs(S)$ is a special EP-EFA without event tags.
\par

\begin{algorithm}
\caption{Construction Algorithm of the Symbolic Observer $Obs(S) = \{ Q^{obs}, q^{obs}_{0}, T^{obs}\}$ for
 $S= (Q, \Sigma, X, Q_{0}, T)$.}
\label{alg:1}
  \uline{\textbf{Step 1)}} Calculate the set of observable transitions $\widehat{T}$ 
  to describe the possible observations to $T$. Specifically, \par
   \par
   \textbf{1.1)} For each \emph{nonzero-step-length}  transition $t = q \xrightarrow[k]{\sigma:\varphi} \widehat{q} \in T$,
    define the predicate for each index set $idx \subseteq [1:k]$ as
   \begin{multline} \label{eq:phi-idx}
     \varphi_{idx} \overset{\text{def}}{=} (\forall j \in [1:k]\backslash idx)(\exists x_{\sigma}^{j})[ \neg \vartheta(x_{\sigma}^{j}) \wedge \bigwedge\nolimits_{i \in idx}\vartheta(x_{\sigma}^{i})  \\
     \wedge \varphi(x_{\sigma}^{1},x_{\sigma}^{2},\ldots, x_{\sigma}^{k})],
   \end{multline}
   where $idx$ is the index set of observable parameters.
   $\varphi_{idx}$ is an $X^{|idx|}$-predicate
   with the free variables $x_{\sigma}^{i}$, $i \in idx$.
\par
   \textbf{1.2)}  For each \emph{nonzero-step-length}  transition
   $t= q \xrightarrow[k]{\sigma:\varphi} \widehat{q}$, construct the set of possible observations $T_t$
   where the element assembles all the possible observable results
    that contain the same number of observable parameters.
 \begin{align} \label{eq:T-t}
 T_{t} = \big \{ q \xrightarrow[i]{\bigvee\nolimits_{|idx| = i}\mathfrak{R}(\varphi_{idx})} \widehat{q} |
    isSat(\bigvee\nolimits_{|idx| = i}\mathfrak{R}(\varphi_{idx})) \nonumber \\
    \wedge i \in [0:k]  \big \},
 \end{align}
 where $\mathfrak{R}$ is a renaming function that changes the names of the free variables in predicates $\varphi_{idx}$  to $x_{\hat{i}}$, $\hat{i} \in [1:i]$, in order.

\par
   \textbf{1.3)} For each \emph{zero-step-length} transition
   $t=q \xrightarrow[0]{\sigma:\varphi} \widehat{q}$,
   \begin{align} \label{eq:T-t2}
     T_{t} =  \{ q \xrightarrow{\epsilon} \widehat{q} \}.
   \end{align}
\par
\textbf{1.4)} Calculate the set of observable transitions
$  \widehat{T} = \bigcup\nolimits_{t \in T} T_{t}$.

\vspace{0.35em}
  \uline{\textbf{Step 2)}} Construct the  observer $Obs(S) = \{ Q^{obs}, q^{obs}_{0}, T^{obs}\}$ by the breadth-first search strategy. Specifically,
   \par
   \textbf{2.1)} Calculate the initial state of observer $q^{obs}_{0}$, which is defined as the following recursive manner
 \begin{equation} \label{eq:qobs0}
 q^{obs}_{0} = Q_{0} \cup \{ \widehat{q} \in Q | q \xrightarrow{\epsilon} \widehat{q}  \in \widehat{T} \wedge q \in q^{obs}_{0}\},
 \end{equation}
 and then put $q^{obs}_{0}$ into the set of states $Q^{obs}$.
\par
 \textbf{2.2)} Do the following for an unvisited state $q^{obs} \in Q^{obs}$,
 until no unvisited state can be found in $Q^{obs}$.
    \begin{itemize}
      \item
    Collect the observable transitions originating from the states in $q^{obs}$,
      and divide them into $k_{m}$ parts, $k_{m} = \max \{ stp(t) | $ $ src(t) \in q^{obs} \wedge t \in \widehat{T} \}$,
      where $stp(t)$ and $src(t)$ are step-length and source state of $t$, respectively.
      Formally, the $k^{th}$, $k \in [1:k_{m}]$, part is defined as follows.
      \begin{equation} \label{eq:tkqobs}
        \widehat{T}^{k}_{q^{obs}} = \{ t \in \widehat{T} | src(t) \in q^{obs} \wedge stp(t) = k \}.
      \end{equation}
      \item
    Index the set $\widehat{T}^{k}_{q^{obs}}$ as
      $\widehat{T}^{k}_{q^{obs}} = \{\hat{t}_{i} | i \in [1: |\widehat{T}^{k}_{q^{obs}}|]\}$.
      Let   $pdc(\hat{t}_{i})$ and $tgt(\hat{t}_{i})$ be the predicate and target state of $\hat{t}_{i}$, respectively.
       For each \emph{nonempty}  set $\widehat{idx} \subseteq [1:|\widehat{T}^{k}_{q^{obs}}|]$,
      \begin{align} \label{eq:psi-idx}
        \psi_{\widehat{idx}} = ( \bigwedge\nolimits_{i \in \widehat{idx}} pdc(\hat{t}_{i}) )    \wedge
        ( \bigwedge\nolimits_{i \notin \widehat{idx}} \neg pdc(\hat{t}_{i})),
      \end{align}
      describes the condition under which only the transitions $\hat{t}_{i}$, $i \in \widehat{idx}$,
      can be fired at $q^{obs}$.
      If $\psi_{\widehat{idx}}$ is satisfiable, let
        \begin{align}
            \widehat{q}^{obs} &= \{ tgt( \hat{t}_{i}) | i \in \widehat{idx} \} \text{, and } \nonumber \\
            \widetilde{q}^{obs} &=  \widehat{q}^{obs} \cup
            \{ \widehat{q} | q \xrightarrow{\epsilon} \widehat{q} \in
       \widehat{T} \wedge q \in \widetilde{q}^{obs} \},   \label{eq:tildeqobs}
        \end{align}
       and then put  $q^{obs} \xrightarrow[k]{\psi_{\widehat{idx}}} \widetilde{q}^{obs}$ into $T^{obs}$,
       and further put $\widetilde{q}^{obs}$ into $Q^{obs}$ if $\widetilde{q}^{obs} \not\in Q^{obs}$.
   \end{itemize}
\end{algorithm}
\par
In the following, we would like to prove that the verification of current-state opacity for the EP-EFA $S$
can be realized by means of its symbolic observer $Obs(S)$.
Firstly, we present three necessary Lemmas.
  \begin{lemma} \label{lemma:est}
   Given an EP-EFA $S =  (Q, \Sigma, X, Q_{0}, T)$ with the set of secret states $Q_{s}$,
    the set of non-secret states $Q_{ns}$, and the observation function $\theta$.
    $S$ is current-state opaque w.r.t. $Q_{s}$, $Q_{ns}$ and $\theta$,
     if and only if for any observation $w \in \Theta(S)$,
     $Est^{S}(w) \cap Q_{s} \neq \varnothing$ $\Rightarrow$ $Est^{S}(w) \cap Q_{ns} \neq \varnothing$.
  \end{lemma}
  \begin{IEEEproof}
  ($\Leftarrow$) Given any $u \in L_{Q_{0}}^{Q_{s}}(S)$.
  Let $w = \theta(\mathfrak{D}(u))$.
  This implies $Est^{S}(w) \cap Q_{s} \neq \varnothing$.
  Thus, we have $Est^{S}(w) \cap Q_{ns} \neq \varnothing$, which
   means that there exist $q_{0} \in Q_{0}$,  a non-secret state $\widehat{q} \in Q_{ns}$
  and a parameterized string $v$, such that $q_{0} \xrightarrow{v} \widehat{q}$,
   and   $w = \theta(\mathfrak{D}(v))$.
  This further implies that $v \in L_{Q_{0}}^{Q_{ns}}(S)$ and $\theta(\mathfrak{D}(u)) = \theta(\mathfrak{D}(v))$.
    According to Definition \ref{def:cso}, $S$ is current-state opaque.
    \par
    ($\Rightarrow$)  Given any observation $w \in \Theta(S)$ satisfying $Est^{S}(w) \cap Q_{s} \neq \varnothing$.
     $Est^{S}(w) \cap Q_{s} \neq \varnothing$  means  there exists a parameterized string $u \in  L_{Q_{0}}^{Q_{s}}(S)$ such that $w = \theta(\mathfrak{D}(u))$.
     Since $S$ is current-state opaque, there exists $v \in L_{Q_{0}}^{Q_{ns}}(S)$ such that
       $\theta(\mathfrak{D}(v)) = \theta(\mathfrak{D}(u)) = w$.
     This means that there exist $q_{0} \in Q_{0}$ and $\widehat{q} \in Q_{ns}$
     such that $q_{0} \xrightarrow{v} \widehat{q}$,
     which implies that $\widehat{q} \in Est^{S}(w)$.
     Thus $Est^{S}(w) \cap Q_{ns} \neq \varnothing$.
     \end{IEEEproof}

\par
  Lemma \ref{lemma:est} implies that the verification of current-state opacity can be realized by
   going through all the possible state estimations.
  The following two Lemmas further prove that the states of the symbolic observer are exactly
  all the state estimations.

\begin{lemma} \label{lemma:obs}
Given an EP-EFA $S =  (Q, \Sigma, X, Q_{0},  T)$ and
its symbolic observer $Obs(S) = \{ Q^{obs}, q^{obs}_{0}, T^{obs} \}$ constructed by Algorithm \ref{alg:1}.
  For any observation $w \in \Theta(S)$,
  $Est^{S}(w)$ is the state reachable from $q^{obs}_{0}$ by $w$ in $Obs(S)$.
\end{lemma}
\begin{IEEEproof}
  Firstly, we claim that $Obs(S)$ constructed by Algorithm \ref{alg:1} is deterministic,
  i.e., given an observation, there exists only one reachable state in $Q^{obs}$.
  This is
  because Equation (\ref{eq:psi-idx}) implies that if $idx1 \neq idx2$, then $\psi_{idx1} \wedge \psi_{idx2} = \bot$, and
  thus no observation unit can simultaneously satisfy two different symbolic transitions originating from the same state
  $q^{obs}$ of $Obs(S)$.

  \par
  Secondly, we prove this Lemma by induction on the number of observation units in $w$.
  Let $|w|_{u} = n$.
  \par
  The base case is $n = 0$, i.e., $w = \epsilon $.
  It is sufficient to show $Est^{S}(\epsilon) = q^{obs}_{0}$.
  If $q \in Q_{0}$, obviously we have $q \in Est^{S}(\epsilon) $ and $ q \in q^{obs}_{0}$.
  The remainder is to show $Est^{S}(\epsilon) \backslash Q_{0}  = q^{obs}_{0} \backslash Q_{0} $.
  According to Equations (\ref{eq:phi-idx}-\ref{eq:T-t2}), $q \xrightarrow{\epsilon} \widehat{q} \in \widehat{T}$
  means there exists a symbolic transition $q \xrightarrow[k]{\sigma:\varphi} \widehat{q}$ in $S$,
  such that $\varphi$ holds for certain $k$ unobservable event parameters or $k=0$.
  Therefore, by Equation (\ref{eq:qobs0}), a state $q_{n} \in q^{obs}_{0}\backslash Q_{0}$,
  if and only if there exist a sequence of
  transitions
  $q_{0} \xrightarrow[k_{1}]{\sigma_{1}:\varphi_{1}} {q_{1}} \ldots \xrightarrow[k_{n}]{\sigma_{n}:\varphi_{n}} {q_{n}}$ in $S$,  $q_{0} \in Q_{0}$,
  where each predicate $\varphi_{i}$ holds for $k_{i}$,  $i \in [1:n]$, unobservable event parameters
  or $k_{i} = 0$.
  This is equivalent to saying that there exists a parameterized string $u$, $\theta(\mathfrak{D}(u)) = \epsilon$, and  $q_{0} \xrightarrow{u} q_{n}$ by Definition \ref{def:longtransition2},
  which also means that $q_{n} \in Est^{S}(\epsilon)$ by Equation (\ref{eq:est}).
  Thus the base case holds.

  \par
  The induction hypothesis is that for all observation $w$, $|w|_{u} \leq n$, $Est^{S}(w)$ is reachable by $w$ in $Obs(S)$.
  We need to show that for any observation unit $\widehat{w} = \langle a_{1}\ldots a_{k}\rangle$, $k \geq 1$,
   such that $w\widehat{w} \in \Theta(S)$,  $Est^{S}(w\widehat{w})$ is reached  by
   $w\widehat{w}$ from $q^{obs}_{0}$ in $Obs(S)$.
   This is equivalent to show that $Est^{S}(w\widehat{w})$ is reachable by $\widehat{w}$ from state $Est^{S}(w)$
   due to the fact that the observer $Obs(S)$ is deterministic.
   Since the observation function $\theta$ is static,
   we can reformulate $Est^{S}(w\widehat{w})$ as follows.
   \begin{equation} \label{eq:estwhatw}
     Est^{S}(w\widehat{w}) = \{  \widehat{q} | q \xrightarrow{\widehat{u}} \widehat{q}  \wedge q \in Est^{S}(w) \wedge
     \widehat{w} = \theta( \mathfrak{D}(\widehat{u})) \}.
   \end{equation}

     \par
     Taking $Est^{S}(w)$ as the $q^{obs}$ in Equation (\ref{eq:tkqobs}),
     then $T^{k}_{Est^{S}(w)}$ is the set of observable transitions that originate from one of the states in $Est^{S}(w)$ and contain $k$ observable parameters.
     Suppose $\widehat{idx} \subseteq [1:|T^{k}_{Est^{S}(w)}|]$ is the \emph{only} nonempty index set such that
     the observation unit $\widehat{w} = \langle a_{1}\ldots a_{k}\rangle$ satisfies $\psi_{\widehat{idx}}$
     (the existence follows from the fact that $w\widehat{w} \in \Theta(S)$ and
     the uniqueness follows from Equation (\ref{eq:psi-idx})).
      According to Equations (\ref{eq:tildeqobs},\ref{eq:estwhatw}),
      we obtain $Est^{S}(w\widehat{w}) = \widetilde{q}^{obs}$.
      By Algorithm \ref{alg:1}, we have $Est^{S}(w) \xrightarrow[k]{\psi_{\widehat{idx}}} \widetilde{q}^{obs} \in T^{obs}$,
      and thus $Est^{S}(w) \xrightarrow[k]{\psi_{\widehat{idx}}} Est^{S}(w\widehat{w}) \in T^{obs}$, which implies
       $Est^{S}(w\widehat{w})$ is reached from $ q^{obs}_{0}$ by $w\widehat{w}$ in $Obs(S)$.
      This completes the proof of the induction step.
         \end{IEEEproof}

\begin{lemma} \label{lemma:obs2}
Given an EP-EFA  $S =  (Q, \Sigma,X, Q_{0},  T)$ and its symbolic observer
$Obs(S) = \{ Q^{obs}, q^{obs}_{0},  T^{obs} \}$.
We have $L(Obs(S)) = \Theta(S)$.
\end{lemma}
\begin{IEEEproof}
 We prove this Lemma by induction on the number of observation units in $w \in L(Obs(S))$.
 Let $|w|_{u} = n$.
  The base case is $n =0$, i.e., $w = \epsilon$. Obviously $\epsilon \in L(Obs(S))$ and
  $\epsilon \in\Theta(S)$. Thus the base case holds.
  \par
  The induction hypothesis is that $w \in L(Obs(S)) \Leftrightarrow w \in \Theta(S)$ holds
  for any observation $w$, $|w|_{u} \leq n$.
  Then we need to show for each observation unit $\widehat{w}=\langle a_{1},\ldots,a_{k}\rangle$,
  $w\widehat{w} \in L(Obs(S)) \Leftrightarrow w\widehat{w} \in \Theta(S)$.
  Suppose $q^{obs}$ is reached by  $w$ from $q^{obs}_{0}$ in $Obs(S)$.
  By Lemma (\ref{lemma:obs}) and Equation (\ref{eq:est}),  for each $q_{i} \in q^{obs}$,
  there exists an initial state $q_{0}^{i} \in Q_{0}$ such that
  $q_{0}^{i} \xrightarrow{u_{i}} q_{i}$,
  $\theta(\mathfrak{D}(u_{i})) = w$.
  By Equations (\ref{eq:psi-idx}-\ref{eq:tildeqobs}), $w\widehat{w} \in L(Obs(S))$ holds if and only if
  there exists a nonempty index set $\widehat{idx}$ such that $\psi_{\widehat{idx}}$ holds for $\widehat{w}$.
  This is equivalent to saying that there exists at least an observable transition
  $(t_{i}=q_{i} \xrightarrow[k]{\varphi_{i}} \widehat{q}_{i}) \in \widehat{T}_{q^{obs}}^{k}$,
  $q_{i} \in q^{obs}$,
   $\widehat{w} \in \llbracket \varphi_{i} \rrbracket$, $i \in \widehat{idx}$,
  which further means there exists a parameterized event $\widehat{u}_{i}$ such that
  $q_{i} \xrightarrow{\widehat{u}_{i}} \widehat{q}_{i}$
  and $\theta(\mathfrak{D}(\widehat{u}_{i})) = \widehat{w}$
  by Equations (\ref{eq:phi-idx}, \ref{eq:T-t}).
  Therefore, $w\widehat{w} \in L(Obs(S))$ holds if and only if there exists $\widehat{u}_{i}$ such that
  $q^{i}_{0} \xrightarrow{u_{i}\widehat{u}_{i}} \widehat{q}_{i}$,
  $\theta(\mathfrak{D}(u_{i})) = w$,
  $\theta(\mathfrak{D}(\widehat{u}_{i})) = \widehat{w}$, which means
  $w\widehat{w} \in \theta(\mathfrak{D}(u_{i}\widehat{u}_{i})) \in \Theta(S)$.
  This completes the proof of the induction step.
   \end{IEEEproof}

\par
Lemma \ref{lemma:obs} implies that the state estimation for each observation is contained in the state space of the
symbolic observer $Obs(S)$.
Lemma \ref{lemma:obs2} further implies that only the observations can reach the states of $Obs(S)$.
Hence, the state space of $Obs(S)$ are \emph{exactly} all the state estimations of $S$.
Therefore, by Lemmas (\ref{lemma:est}, \ref{lemma:obs}, \ref{lemma:obs2}), we have the following theorem.
\par

\begin{theorem} \label{th:cso}
  Given an EP-EFA $S =  (Q, \Sigma, X, Q_{0}, T)$ with the set of secret states $Q_{s}$,
   the set of non-secret states $Q_{ns}$, and the observation function $\theta$.
  Let $Obs(S) = \{ Q^{obs}, $ $ q^{obs}_{0},  T^{obs}\}$ be the symbolic observer
  constructed by Algorithm \ref{alg:1}.
  $S$ is current-state opaque w.r.t. $Q_{s}$, $Q_{ns}$ and $\theta$,
  if and only if  for any $q^{obs} \in Q^{obs}$,
  \begin{equation} \label{eq:cso2}
  q^{obs} \cap Q_{s} \neq \varnothing \Rightarrow q^{obs} \cap Q_{ns} \neq \varnothing.
  \end{equation}
\end{theorem}

\begin{remark}
 The verification of current-state opacity and the construction of the symbolic observer (Algorithm \ref{alg:1}) have the same complexity,
as  checking the validness for Equation (\ref{eq:cso2}) can be finished during the construction of $Obs(S)$.
 Suppose  the EP-EFA $S=  (Q, \Sigma, X, Q_{0}, T)$ with $K$ step-length has $N$ states
 and $M$ symbolic transitions.
  Assume that $g(z)$ is the cost of checking satisfiability of the predicate with $z$ free variables in the Boolean algebra.
  In Step 1) of Algorithm \ref{alg:1}, we have $|\widehat{T}| \leq M*(K+1)$, and for each symbolic transition with $l$ step-length, $l+1$ predicates are checked for satisfiability. Thus, the complexity of Step 1) is at most $M*(K+1)*g(K)$.
  In Step 2), for each $\widehat{T}^{k}_{q^{obs}}$, there are $2^{|\widehat{T}^{k}_{q^{obs}}|}-1$ combined predicates that need to be checked for satisfiability.
  Hence, there are at most $\sum_{q^{obs} \in Q^{obs}}\sum_{k=1}^{K}(2^{|T^{k}_{q^{obs}}|}-1)$ predicates are checked for satisfiability.
  For a given $q^{obs}$,
   we have $\sum_{k=1}^{K}|\widehat{T}^{k}_{q^{obs}}| \leq |\widehat{T}|$, and by this equation,
  we can prove  $\sum_{k=1}^{K} (2^{|T^{k}_{q^{obs}}|}-1) <  2^{|\widehat{T}|} \leq 2^{M*(K+1)}$.
  Since $|Q^{obs}| \leq 2^{N}$,  the complexity of Step 2) of Algorithm \ref{alg:1} is at most $g(K) * 2^{N}*2^{M*(K+1)}$.
  Therefore, the complexity of the verification of current-state opacity is  $g(K) * 2^{N+M*K}$.
\end{remark}

\par
As aforementioned, the EP-EFAs model can address many complex data and operations via the symbolic transitions.
However, for the simplicity to demonstrate the obtained results,
the following illustrative examples only consider integer arithmetic.
\par

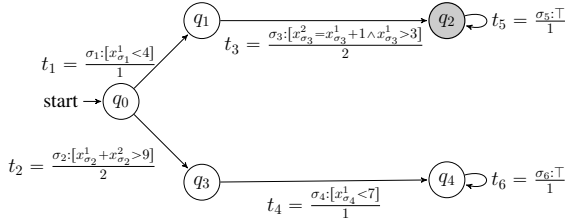
\begin{figure}[h]
\centering
    \resizebox{0.42\textwidth}{!}{%
     \begin{tikzpicture}[font=\Large,->,>=stealth',shorten >=1pt,auto,node distance=7.5cm,semithick,
                                  every state/.style={font=\Large},
                                  new-qs/.style={fill=gray!40,thick,font=\Large}]
                \node[initial, state] (q0) {$ q_{0} $};
                \node[state] [node distance=2.8cm] [above right of = q0] (q1) {$q_{1} $};
                \node[state,new-qs] [node distance=6.0cm] [right of = q1] (q2) {$q_{2} $};
                \node[state] [node distance=3.9cm] [below of = q2] (q4) {$q_{4} $};
                \node[state] [node distance=2.8cm] [below right of = q0] (q3) {$q_{3} $};

            \draw[every node/.style={font=\Large}]
     (q0)[left]  edge  node{$t_{1}=\frac{\sigma_{1}:[x_{\sigma_{1}}^{1} < 4]}{1} $} (q1)
       (q0)[below left]  edge  node{$t_{2}=\frac{\sigma_{2}:[x_{\sigma_{2}}^{1} + x_{\sigma_{2}}^{2} > 9]}{2}$} (q3)
      (q1)[below] edge  node{$t_{3}=\frac{\sigma_{3}: [x_{\sigma_{3}}^{2} = x_{\sigma_{3}}^{1} +1 \wedge x_{\sigma_{3}}^{1} > 3]}{2} $} (q2)
       (q3)  [below] edge  node{$t_{4}=\frac{\sigma_{4}: [x_{\sigma_{4}}^{1}<7]}{1}$} (q4)
       (q2) edge[loop right] node{$t_{5}=\frac{\sigma_{5}:\top}{1}$} ( )
       (q4) edge[loop right] node{$t_{6}=\frac{\sigma_{6}:\top}{1}$} ( );
               \end{tikzpicture}
            }
\caption{The EP-EFA $S$ with $X = \mathbb{N}$ in Example \ref{ex:3}}
\label{fig-7}
\end{figure}

\begin{figure}[h]
\centering
    \resizebox{0.44\textwidth}{!}{%
     \begin{tikzpicture}[font=\Large,->,>=stealth',shorten >=1pt,auto,semithick,
                                  every state/.style={rectangle,minimum width=0.12\textwidth, font=\Large},
                                  new-qs/.style={fill=gray!40,thick,font=\Large}]
                \node[initial, state] (q0) {$ q_{0},q_{1} $};
                \node[state,new-qs] [node distance=3cm] [above of = q0] (q1) {$q_{2}$};
                \node[state,new-qs] [node distance=6.5cm] [right of = q0] (q2) {$q_{2},q_{3},q_{4} $};
                \node[state] [node distance=3cm] [below of = q0] (q3) {$q_{3},q_{4} $};
                \node[state] [node distance=3cm] [below of = q2] (q4) {$q_{4} $};
                \node[state,new-qs] [node distance=3cm] [above of = q2] (q5) {$q_{2},q_{4}$};

            \draw[every node/.style={font=\Large}]
     (q0)[left]  edge  node{$\frac{[x_{1} = 5]}{1} $} (q1)
       (q0) [left]  edge  node{$\frac{[x_{1}>5]}{1}$} (q3)
      (q0) [below] edge  node{$\frac{[x_{1} \geq 5 \wedge x_{2} \geq 5 \wedge x_{2} = x_{1} +1]}{2} $} (q2)
       (q3)  [below] edge  node{$\frac{[ x_{1} \geq 5]}{1}$} (q4)
       (q2)  [right] edge  node{$\frac{[ x_{1} \geq 5]}{1}$} (q5)
       (q1) edge[loop right] node{$\frac{[x_{1} \geq 5]}{1}$} ( )
       (q4) edge[loop right] node{$\frac{[x_{1} \geq 5]}{1}$} ( )
       (q5) edge[loop right] node{$\frac{[x_{1} \geq 5]}{1}$} ( )
       (q0) [right] [bend left]  edge  node{$\frac{[x_{1} \geq 5 \wedge x_{2} \geq 5 \wedge x_{2} \neq x_{1} +1]}{2}$} (q3);
               \end{tikzpicture}
            }
\caption{The symbolic observer of the EP-EFA $S$ obtained by Algorithm \ref{alg:1}.}
\label{fig-8}
\end{figure}
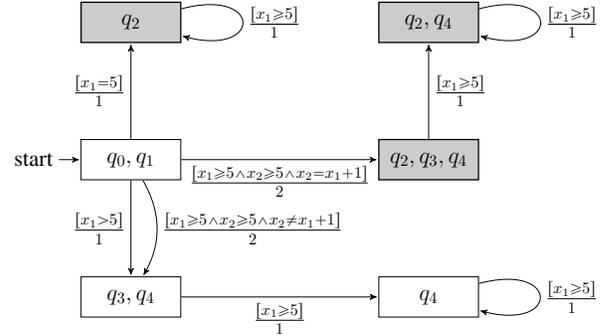

\begin{example} \label{ex:3}
  Consider an EP-EFA  $S$ with $X = \mathbb{N}$ shown in Fig. \ref{fig-7},
  where the set of secret states $Q_{s} = \{q_{2}\}$ and
  the set of non-secret states $Q_{ns} = Q\backslash Q_{s}$.
  Suppose that the observation function $\theta$ is obtained by the  $X$-predicate
  $\vartheta(x) \overset{\text{def}}{=} [x \geq 5 ]$.
  \par
  Firstly, we construct the observable transitions as follows.
\begin{align*}
  T_{t_{1}}= \{&
  q_{0} \xrightarrow{\epsilon} q_{1}
  \}.
  \\
T_{t_{2}}= \{ &
   q_{0} \xrightarrow[2]{x_{1} \geq 5 \wedge x_{2} \geq 5} q_{3};   q_{0} \xrightarrow[1]{
  (\exists x_{2} <5) x_{1} + x_{2} > 9  \wedge x_{1} \geq 5} q_{3}
  \}.
\\
  T_{t_{3}}= \{ &
  q_{1} \xrightarrow[1]{x_{1} = 5 } q_{2} ; \quad
  q_{1} \xrightarrow[2]{x_{1} \geq 5 \wedge x_{2} = x_{1} + 1 } q_{2}
  \}.
\\
T_{t_{4}}= \{ &
  q_{3} \xrightarrow{\epsilon} q_{4} ; \quad
   q_{3} \xrightarrow[1]{x_{1} <7 \wedge x_{1} \geq 5} q_{4}
  \}.
  \\
T_{t_{5}}= \{ &
  q_{2} \xrightarrow{\epsilon} q_{2} ; \quad
   q_{2} \xrightarrow[1]{ x_{1} \geq 5} q_{2}
  \}.
  \\
T_{t_{6}}= \{ &
   q_{4} \xrightarrow{\epsilon} q_{4} ; \quad
   q_{4} \xrightarrow[1]{ x_{1} \geq 5} q_{4}
  \}.
  \end{align*}
\par
Secondly, we have $q^{obs}_{0} = \{q_{0}, q_{1}\}$, and obtain the corresponding set as follows.
        \[
           \widehat{T}^{2}_{q^{obs}_{0}} =
             \{
             q_{0} \xrightarrow[2]{x_{1} \geq 5 \wedge x_{2} \geq 5} q_{3} ; \quad
             q_{1} \xrightarrow[2]{x_{1} \geq 5 \wedge x_{2} = x_{1} + 1 } q_{2}
               \}.
           \]          \[
           \widehat{T}^{1}_{q^{obs}_{0}} =
             \{
             q_{1}\xrightarrow[1]{x_{1} = 5 } q_{2}; \quad
             q_{0} \xrightarrow[1]{(\exists x_{2} <5) x_{1} + x_{2} > 9  \wedge x_{1} \geq 5} q_{3}
               \}.
           \]

\par
For $\widehat{T}^{2}_{q^{obs}_{0}}$, the set of satisfiable combined predicates are as follows.
    \begin{subequations}\label{ex3-3}
        \begin{align*}
            \Psi(\widehat{T}^{2}_{q^{obs}_{0}}) = \{
            \psi_{\{1\}} = [x_{1} \geq 5 \wedge x_{2} \geq 5 \wedge x_{2} \neq x_{1} +1];    \nonumber \\
            \psi_{\{1,2\}} = [x_{1} \geq 5 \wedge x_{2} \geq 5 \wedge x_{2} = x_{1} +1]
            \}.
        \end{align*}
    \end{subequations}
Through the transitions guarded with $\psi_{\{1\}}$ and $\psi_{\{1,2\}}$,
the states $\{ q_{3}, q_{4} \}$ and $\{q_{2}, q_{3}, q_{4}\}$ are, respectively,
 generated and  put into $Q^{obs}$.
In addition, the corresponding symbolic transitions are put into $T^{obs}$.
For $\widehat{T}^{1}_{q^{obs}_{0}}$,  the set of satisfiable combined predicates are as follows.
   \begin{subequations}\label{ex3-4}
        \begin{align*}
            \Psi(\widehat{T}^{1}_{q^{obs}_{0}}) = \{
            \psi_{\{1\}} = [x_{1} = 5 ]; \quad
            \psi_{\{2\}} = [x_{1} > 5 ]\}.
        \end{align*}
    \end{subequations}
Through the transitions guarded with $\psi_{\{1\}}$ and $\psi_{\{2\}}$,
the states $\{ q_{2} \}$, $\{ q_{3}, q_{4} \}$ are, respectively, generated
and the former is put into $Q^{obs}$.
Meanwhile, the corresponding symbolic transitions are put into $T^{obs}$.

\par
 For other unvisited states in $Q^{obs}$, we do the same things as that for state $q^{obs}_{0}$.
 Finally, we obtain the symbolic observer, as shown in Fig. \ref{fig-8}, where
 $Q^{obs} = \{ \{ q_{0},q_{1}\}, \{q_{2}\}, \{ q_{3},q_{4}\}, $ $ \{ q_{2},q_{3},q_{4}\},
  \{q_{4}\}, \{ q_{2},q_{4}\}  \}$.
 For the state $\{q_{2}\}  \in Q^{obs}$,
 we have $\{ q_{2} \} \cap Q_{s} \neq \varnothing$ and $\{ q_{2} \} \cap Q_{ns} = \varnothing$.
 Hence by Theorem \ref{th:cso}, $S$ is not current-state opaque w.r.t. $\{q_{2}\}$, $\{q_{0},q_{1},q_{3},q_{4}\}$
  and $\theta$.
\end{example}

\subsection{Initial-State Opacity of EP-EFAs}

The coding manners of the secret behavior in current-state opacity and initial-state opacity are  reverse.
According to this property, we transform the verification of initial-state opacity into
the verification of current-state opacity for EP-EFAs.

\par
Firstly, we define the reverse operations for parameterized strings, data strings and symbolic transitions.
Given a parameterized string
$u = \sigma_{1}\langle a_{{1}}^{1}, a_{{1}}^{2}, \ldots, a_{{1}}^{k_{1}}\rangle$
 $\sigma_{2}\langle a_{{2}}^{1}, a_{{2}}^{2}, \ldots, a_{{2}}^{k_{2}}\rangle$
 $ \ldots$
  $ \sigma_{n}\langle a_{{n}}^{1}, a_{{n}}^{2}, \ldots, a_{{n}}^{k_{n}}\rangle$,
 its \emph{reverse} is
 $u^{r} \overset{\text{def}}{=} \sigma_{n}\langle a_{{n}}^{k_{n}}, \ldots, a_{{n}}^{2}, a_{{n}}^{1}\rangle$
 $ \ldots$
 $\sigma_{2}\langle a_{{2}}^{k_{2}}, \ldots, a_{{2}}^{2},a_{{2}}^{1}\rangle$
  $\sigma_{1}\langle a_{{1}}^{k_{1}}, \ldots, a_{{1}}^{2}, a_{{1}}^{1}\rangle$.
For a data string $d = \mathfrak{D}(u)$,
the \emph{reverse} of $d$ is
$d^{r} \overset{\text{def}}{=}\mathfrak{D}(u^{r})$.
For a symbolic  transition $t =  q \xrightarrow[k]{\sigma:\varphi} \widehat{q}$,
the \emph{reverse} of $t$ is defined as
$t^{r} \overset{\text{def}}{=} \widehat{q} \xrightarrow[k]{\sigma:\varphi^{r}} q$,
where the predicate $\varphi^{r}$ is obtained from
$\varphi$ by changing the name of the free variable $x^{i}_{\sigma}$ to $x^{k+1 -i}_{\sigma}$, $i \in [1:k]$,
e.g., the reverse of the $X^{4}$-predicate
$\varphi = [x^{1}_{\sigma} > x^{3}_{\sigma} \wedge x^{2}_{\sigma} \neq x^{4}_{\sigma}]$ is
$\varphi^{r} \overset{\text{def}}{=} [x^{4}_{\sigma} > x^{2}_{\sigma} \wedge x^{3}_{\sigma} \neq x^{1}_{\sigma}]$.
By the aforementioned definitions, we have $d \in \llbracket \varphi  \rrbracket$ if and only if
$d^{r} \in \llbracket \varphi^{r}  \rrbracket$.
\par
\begin{definition} \label{def:rev}
  Given an EP-EFA $S =  (Q, \Sigma, X, Q_{0}, T)$.
  The reverse of $S$ is defined as
  $S^{r} =  (Q, \Sigma, X, Q_{0}^{r}, T^{r})$, where the set of initial states is $Q_{0}^{r} = Q$  and
   the set of symbolic transitions is $T^{r} = \{t^{r}|t\in T\}$.
\end{definition}

\par
Definition \ref{def:rev} generalizes the notion of reverse automata \cite{book-computation},
which has been widely used in many fields.
In particular,
by constructing the observer for  reverse finite automata, Wu \emph{et al.} \cite{ifo} proposed an approach to verify
the initial-state opacity for classic DESs.
 \par

\par
The following proposition  follows from the definitions of the reverse operations, symbolic transitions,  languages and observations.
\par
\begin{proposition} \label{pro:reverse}
Given a transition $t$, a parameter tuple $d$, a parameterized string $u$, an observation $w$, and
an EP-EFA $S =  (Q, \Sigma, X, Q_{0},  T)$ and its reverse $S^{r} =  (Q, \Sigma, X, Q_{0}^{r}, T^{r})$.
The following equations hold. \par
1) $d \in \llbracket prd(t)  \rrbracket  \Longleftrightarrow d^{r} \in \llbracket prd(t^{r})  \rrbracket$,
   where the $pdc(t)$ and $pdc(t^{r})$ denote the predicates of $t$ and $t^{r}$, respectively. \par
2) $q \xrightarrow{u} \widehat{q}  \Longleftrightarrow\widehat{q} \xrightarrow{u^{r}} q $. \par
3) $u \in L_{Q_{1}}^{Q_{2}}(S)  \Longleftrightarrow u^{r} \in L_{Q_{2}}^{Q_{1}}(S^{r})$. \par
4) $w = \theta(\mathfrak{D}(u)) \Longleftrightarrow  w^{r} = \theta(\mathfrak{D}(u^{r}))$. \par
\end{proposition}

\begin{theorem} \label{thm:isotocso}
 Given an EP-EFA  $S =  (Q, \Sigma, X, Q_{0}, T)$ with the set of secret initial states $Q_{s} \subseteq Q_{0}$,
  the set of non-secret initial states $Q_{ns} \subseteq Q_{0}$,
  and  observation function $\theta$.
   The reverse of $S$ is $S^{r} = (Q, \Sigma, X, Q_{0}^{r}, T^{r})$ where $Q_{0}^{r} = Q$.
   $S$ is initial-state opaque w.r.t. $Q_{s}$, $Q_{ns}$ and $\theta$,
   if and only if $S^{r}$ is current-state opaque w.r.t. $Q_{s}$, $Q_{ns}$ and $\theta$.
\end{theorem}
\begin{IEEEproof}
By Definition \ref{def:iso},  $S$ is initial-state opaque w.r.t. $Q_{s}$, $Q_{ns}$ and $\theta$,  if and only if
$(\forall u \in L^{Q}_{Q_{s}}(S)) (\exists v \in L^{Q}_{Q_{ns}}(S)) $ $ \theta(\mathfrak{D}(u))=\theta(\mathfrak{D}(v))$.
This is equivalent to
$(\forall u^{r} \in L^{Q_{s}}_{Q}(S^{r})) $ $ (\exists v^{r} \in L^{Q_{ns}}_{Q}(S^{r}))  \theta(\mathfrak{D}(u^{r})) = \theta(\mathfrak{D}(v^{r})) $ by  Proposition \ref{pro:reverse}.
This means $S^{r}$ is current-state opaque w.r.t. $Q_{s}$, $Q_{ns}$ and $\theta$ according to Definition \ref{def:cso}.
   \end{IEEEproof}

\begin{remark}
Theorem \ref{thm:isotocso} implies that the verification of initial-state opacity can be efficiently reduced to
the verification of current-state opacity.
Since the reverse SPA-EFA $S^{r}$ has the same scale as $S$,
the complexity of the verification of initial-state opacity  is also $g(K) * 2^{N+M*K}$.
\end{remark}

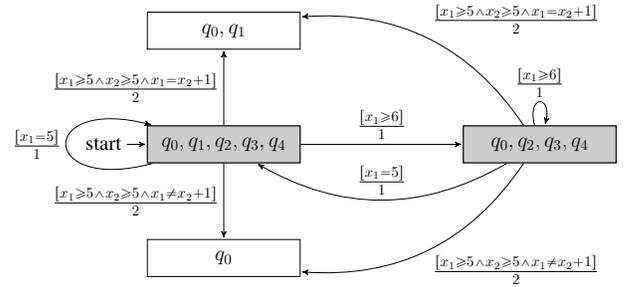
\begin{figure}[h]
\centering
    \resizebox{0.45\textwidth}{!}{%
     \begin{tikzpicture}[font=\Large,->,>=stealth',shorten >=1pt,auto,semithick,
                                  every state/.style={rectangle,minimum width=0.2\textwidth, font=\Large},
                                  new-qs/.style={fill=gray!40,thick,font=\Large}]
                \node[initial, state,new-qs] (q0) {$ q_{0},q_{1},q_{2},q_{3},q_{4}$};
                \node[state,new-qs] [node distance=7.5cm] [ right of = q0] (q1) {$q_{0},q_{2},q_{3},q_{4}$};
                \node[state] [node distance=2.7cm] [above of = q0] (q2) {$q_{0},q_{1} $};
                \node[state] [node distance=2.7cm] [below of = q0] (q3) {$q_{0}$};

            \draw[every node/.style={font=\Large}]
(q0) [left] edge  node{$\frac{[x_{1} \geq 5 \wedge x_{2} \geq 5 \wedge x_{1} = x_{2} +1]}{2} $} (q2)
(q0) [left]   edge  node{$\frac{[x_{1} \geq 5 \wedge x_{2} \geq 5 \wedge x_{1} \neq x_{2} +1]}{2}$} (q3)
(q1) edge[loop above] node{$\frac{[x_{1} \geq 6]}{1}$} ( )
(q0) edge[loop  left] node{$\frac{[x_{1} = 5]}{1}$} ( )
(q0)[above]   edge  node{$\frac{[x_{1} \geq 6]}{1} $} (q1)
(q1)[above]  [bend left] edge  node{$\frac{[x_{1} = 5]}{1} $} (q0)
(q1) [above right] [bend right] edge  node{$\frac{[x_{1} \geq 5 \wedge x_{2} \geq 5 \wedge x_{1} = x_{2} +1]}{2} $} (q2)
(q1) [below right]  [bend left]  edge  node{$\frac{[x_{1} \geq 5 \wedge x_{2} \geq 5 \wedge x_{1} \neq x_{2} +1]}{2}$} (q3);
               \end{tikzpicture}
            }
\caption{The symbolic observer $Obs(S^{r})$ of  $S^{r}$ obtained by Algorithm \ref{alg:1}.}
\label{fig-9}
\end{figure}

\begin{example} \label{ex:4}
  Consider the EP-EFA $S$ shown in Fig. \ref{fig-8} with the same observation function $\theta$
  as that in Example \ref{ex:3}.
  Suppose the set of initial states is $Q_{0} = \{q_{0}, q_{1}, q_{2}\}$,
  and the secret initial states and non-secret initial states are
   $Q_{s} = \{ q_{2} \}$ and $Q_{ns} = \{ q_{0}, q_{1}\}$, respectively.
   \par
   For the reverse SPA-EFA $S^{r} = (Q, \Sigma, X, Q, T^{r})$,
   we construct the symbolic observer $Obs(S^{r}) = \{ Q^{obs}_{r}, q^{obs}_{0}, T^{obs}_{r} \}$ according to Algorithm \ref{alg:1}.
   For the initial state of the observer $q^{obs}_{0} = Q$,
   we obtain the subsets of observable transitions as follows.
     \begin{align*}
           \widehat{T}^{2}_{q^{obs}_{0}} =
             \{
             q_{3} \xrightarrow[2]{x_{1} \geq 5 \wedge x_{2} \geq 5} q_{0} ; \quad
             q_{2} \xrightarrow[2]{ x_{1} = x_{2} + 1 \wedge x_{2} \geq 5} q_{1}
               \}.  \\
           \widehat{T}^{1}_{q^{obs}_{0}} =
             \{
             q_{2}\xrightarrow[1]{x_{1} = 5 } q_{1}; \quad
             q_{3} \xrightarrow[1]{(\exists x_{2} <5) x_{1} + x_{2} > 9  \wedge x_{1} \geq 5} q_{0};  \quad  \nonumber \\
             q_{4} \xrightarrow[1]{x_{1} <7 \wedge x_{1} \geq 5} q_{3}; \quad q_{2} \xrightarrow[1]{x_{1} \geq 5} q_{2};
            \quad q_{4} \xrightarrow[1]{x_{1} \geq 5 } q_{4}
               \}.
     \end{align*}
     \par
  For $\widehat{T}^{2}_{q^{obs}_{0}}$, the set of satisfiable combined predicates are as follows.
        \begin{align*}
            \Psi(\widehat{T}^{2}_{q^{obs}_{0}}) = \{
            \psi_{\{1\}} = [x_{1} \geq 5 \wedge x_{2} \geq 5 \wedge x_{1} \neq x_{2} +1];  \nonumber \\
            \psi_{\{1,2\}} = [x_{1} \geq 5 \wedge x_{2} \geq 5 \wedge x_{1} = x_{2} +1]
            \}.
        \end{align*}
Through the transitions guarded with the above predicates,  
the states $\{ q_{0} \}$ and $\{q_{0}, q_{1}\}$ are generated and put into $Q^{obs}_{r}$.
   For $\widehat{T}^{1}_{q^{obs}_{0}}$,  the set of satisfiable combined predicates are as follows.
        \begin{align*}
            \Psi(\widehat{T}^{1}_{q^{obs}_{0}}) =& \{  \psi_{\{1,3,4,5\}} = [x_{1} = 5 ]; \psi_{\{2,3,4,5\}} = [x_{1} = 6 ];\\
           & \psi_{\{2,4,5\}} = [x_{1} \geq 7 ];
            \}.
        \end{align*}
  Through the transitions guarded with the above predicates,
the states $\{ q_{0}, q_{1}, q_{2}, q_{3},q_{4} \}$ and $\{ q_{0}, q_{2}, q_{3}, q_{4} \}$ are generated
and put into $Q^{obs}_{r}$.

\par
Similarly, we handle other unvisited states in $Q^{obs}_{r}$,
and obtain the symbolic observer $Obs(S^{r})$, shown in Fig. \ref{fig-9}, where
$Q^{obs}_{r} = \big \{ \{q_{0},q_{1},q_{2},q_{3},q_{4}\}, \{q_{0},q_{2},q_{3},q_{4}\}, \{q_{0},q_{1}\}, \{q_{0}\} \big \}$.
 Notice that for each state $q^{obs}_{r} \in Q^{obs}_{r}$,
 $q^{obs}_{r} \cap Q_{s} \neq \varnothing$ always implies $q^{obs}_{r} \cap Q_{ns} \neq \varnothing$,
 thus $S^{r}$ is  current-state opaque  w.r.t. $\{q_{2}\}$, $\{q_{0},q_{1}\}$ and $\theta$.
 By Theorem \ref{thm:isotocso}, $S$ is  initial-state opaque  w.r.t. $\{q_{2}\}$, $\{q_{0},q_{1}\}$ and $\theta$.
\end{example}

\subsection{Infinite-Step Opacity of EP-EFAs}

Yin \emph{et al.} \cite{infinite} presented an ingenious method to verify the infinite-step opacity of FAs
by combining the observers of the obverse and reverse automata (called two-way observers in \cite{infinite}).
Following this idea, we have the theorem as follows.
\par

\begin{theorem} \label{theorem:infinite-spf}
 Given an EP-EFA  $S =  (Q, \Sigma, X, Q_{0}, T)$ with the set of secret states $Q_{s}$,
  the set of non-secret  states $Q_{ns}$,
   and the observation function $\theta$.
   The reverse of $S$ is $S^{r} = (Q, \Sigma, X, Q_{0}^{r}, T^{r})$ where $Q_{0}^{r} = Q$.
   $S$ is infinite-step opaque w.r.t. $Q_{s}$, $Q_{ns}$ and $\theta$,
   if and only if
   \begin{align} \label{eq:infinite-spf}
    &  (\forall w \in \Theta(S))(\forall \widehat{w}^{r} \in \Theta(S^{r}))
     [Est^{S}(w) \cap Est^{S^{r}}(\widehat{w}^{r}) \cap Q_{s}  \nonumber \\
    &  \neq \varnothing \Rightarrow Est^{S}(w) \cap Est^{S^{r}}(\widehat{w}^{r}) \cap Q_{ns} \neq \varnothing].
   \end{align}
\end{theorem}
\begin{IEEEproof}
By Equation (\ref{eq:est}), $Est^{S}(w)$ and $Est^{S^{r}}(\widehat{w}^{r})$ are
$\{\widehat{q} \in Q | q_{0} \xrightarrow{u} \widehat{q}  \wedge q_{0} \in Q_{0} \wedge w = \theta(\mathfrak{D}(u))\}$ and
$\{\widehat{q} \in Q | q \xrightarrow{\widehat{u}^{r}} \widehat{q}  \wedge q \in Q \wedge \widehat{w}^{r} = \theta(\mathfrak{D}(\widehat{u}^{r}))\}$, respectively, and the latter further implies that
$\{\widehat{q} \in Q | \widehat{q} \xrightarrow{\widehat{u}} q  \wedge q \in Q \wedge \widehat{w} = \theta(\mathfrak{D}(\widehat{u}))\}$ by  Proposition \ref{pro:reverse}.
Therefore, $Est^{S}(w) \cap Est^{S^{r}}(\widehat{w}^{r}) \cap Q^{s}$ and
$Est^{S}(w) \cap Est^{S^{r}}(\widehat{w}^{r}) \cap Q^{ns}$, respectively, are equivalent to
\begin{subequations}
\begin{align}
A=\{& \widehat{q} \in Q_{s} |  q_{0} \xrightarrow{u} \widehat{q}  \wedge \widehat{q} \xrightarrow{\widehat{u}} q  \wedge q_{0} \in Q_{0} \wedge  \nonumber \\
 &  w = \theta(\mathfrak{D}(u))  \wedge \widehat{w} = \theta(\mathfrak{D}(\widehat{u})) \}, \text{ and } \label{eq:A}  \\
B=\{& \widehat{q} \in Q_{ns} |  q_{0} \xrightarrow{v} \widehat{q}  \wedge \widehat{q} \xrightarrow{\widehat{v}} q  \wedge q_{0} \in Q_{0} \wedge  \nonumber \\
 & w = \theta(\mathfrak{D}(v))  \wedge \widehat{w} = \theta(\mathfrak{D}(\widehat{v})) \}. \label{eq:B}
\end{align}
\end{subequations}
\par
To complete the proof, it is sufficient to show the equivalence between Equations (\ref{eq:infinite}) and (\ref{eq:infinite-spf}).
Firstly, we prove that Equation (\ref{eq:infinite-spf}) implies Equation (\ref{eq:infinite}).
For any $u\widehat{u} \in L(S)$ satisfying $u \in L_{Q_{0}}^{Q_{s}}(S)$,
let $w_{1} = \theta(\mathfrak{D}(u))$, $\widehat{w_{1}} = \theta(\mathfrak{D}(\widehat{u}))$,
and then we have $w_{1} \in \Theta(S)$ and $\widehat{w_{1}}^{r} \in \Theta(\widehat{S}^{r})$.
Taking the $w_{1}$ and $\widehat{w_{1}}$ here as the $w$ and $\widehat{w}$ in Equation (\ref{eq:A}),
then we have $A \neq \varnothing$.
By Equation (\ref{eq:infinite-spf}), we have $B \neq \varnothing$.
which implies that there exists $v\widehat{v} \in L(S)$ satisfying $\widehat{v} \in L_{Q_{0}}^{Q_{ns}}(S)$,
such that $\theta(\mathfrak{D}(u))  = \theta(\mathfrak{D}(v))$
and $\theta(\mathfrak{D}(\widehat{u})) = \theta(\mathfrak{D}(\widehat{v}))$.
This means that Equation (\ref{eq:infinite}) holds.
\par
Secondly, we prove Equation (\ref{eq:infinite}) implies Equation (\ref{eq:infinite-spf}).
For any $w \in \Theta(S)$ and $\widehat{w}^{r} \in \Theta(S^{r})$ satisfying $A \neq \varnothing$,
we have $u\widehat{u} \in L(S)$, such that $u \in L^{Q_{s}}_{Q_{0}}(S)$,
$w = \theta(\mathfrak{D}(u))$ and  $\widehat{w} = \theta(\mathfrak{D}(\widehat{u}))$.
By Equation (\ref{eq:infinite}), there exists
$ v\widehat{v} \in L(S)$ such that $v \in L^{Q_{ns}}_{Q_{0}}(S)$,
$\theta(\mathfrak{D}(u))=\theta(\mathfrak{D}(v))$ and
$\theta(\mathfrak{D}(\widehat{u}))=\theta(\mathfrak{D}(\widehat{v}))$, which implies $B \neq \varnothing$.
Therefore, Equation (\ref{eq:infinite-spf}) holds.
\end{IEEEproof}

\par
According to Lemmas \ref{lemma:obs}, \ref{lemma:obs2}, the state space of the observer of an EP-EFA
 are exactly the set of state estimations.
By Theorem \ref{theorem:infinite-spf}, the verification of infinite-step opacity can be realized
by going through the state spaces of $Obs(S)$ and $Obs(S^{r})$.
Hence, we have the following algorithm (Algorithm \ref{alg:2}) to verify the infinite-step opacity of EP-EFAs.
\par

\begin{algorithm}
\caption{ Infinite-step Opacity Verification Algorithm of a Given EP-EFA $S=  (Q, \Sigma, X, Q_{0}, T)$}
\label{alg:2}
\uline{\textbf{Step 1)}} Calculate the symbolic observer $Obs(S) = \{ Q^{obs}, $ $ q^{obs}_{0}, T^{obs}\}$ for $S$
 by Algorithm \ref{alg:1}. \par
\uline{\textbf{Step 2)}} Calculate the symbolic observer $Obs(S^{r}) = \{ Q^{obs}_{r}, $ $ Q, T^{obs}_{r}\}$ for $S^{r}$
by Algorithm \ref{alg:1}. \par
\uline{\textbf{Step 3)}} For each pair of states $(q^{obs}, q^{obs}_{r}) \in Q^{obs} \times Q^{obs}_{r}$
satisfying $q^{obs} \cap q^{obs}_{r} \cap Q_{s} \neq \varnothing$,
 check whether $q^{obs} \cap q^{obs}_{r} \cap Q_{ns} \neq \varnothing$ holds or not.
 If the answers are positive for all the pairs of states, then $S$ is infinite-step opaque,
 otherwise $S$ is not infinite-step opaque.
\end{algorithm}

\par
\begin{remark}
  As discussed before, the complexity of step 1) and step 2) of Algorithm \ref{alg:2}
  is $g(K) * 2^{N+M*K}$.
  Since $|Q^{obs}| \leq 2^{N}$ and $|Q^{obs}_{r}| \leq 2^{N}$, the complexity of Step 3)
   of Algorithm \ref{alg:2} is $4^{N}$.
  Therefore,  the complexity of the verification of infinite-step opacity
  is $g(K) * 2^{N+M*K} + 4^{N}$.
\end{remark}
\begin{example}
  Consider the EP-EFA shown in Fig. \ref{fig-7}, where the set of secret states
  $Q_{s} =\{ q_{3} \}$ and the set of non-secret states $Q_{ns} = \{ q_{4} \}$.
  The $Obs(S)$ and $Obs(S^{r})$ have been calculated in Examples \ref{ex:3} and \ref{ex:4},
  as shown in Fig. \ref{fig-8}  and Fig. \ref{fig-9}, respectively.
  Notice that $q^{obs} \cap q^{obs}_{r} \cap \{q_{3}\} \neq \varnothing$ implies
  $q^{obs} \cap q^{obs}_{r} \cap \{q_{4}\} \neq \varnothing$ for all  the pairs of states
  $(q^{obs},q^{obs}_{r}) \in Q^{obs} \times Q^{obs}_{r}$.
  Therefore, $S$ is infinite-step opaque w.r.t. $\{q_{3}\}$, $\{q_{4}\}$ and $\theta$.
\end{example}

\section{Conclusion}
In this paper, we have investigated two parametric DESs models, i.e., EFAs and EP-EFAs, and then established a preliminary opacity theory for parametric DESs, which lays a  foundation to analyze the opacity for complex systems.
The parametric DESs well extends the classic DESs by means of the symbolic transitions
 carrying predicates  over the infinite parameter space.
 The parametric DESs can efficiently represent and process many real-world data with the help of SMT solvers.
It has been illustrated that the coexistence of state and event parameters in the predicates
not only enhances the parametric  model but also complicates it.
Specifically,
we have proved that EFAs model is more expressive than EP-EFAs model,
and also proved that the opacity properties of EFAs are  undecidable in general.
In addition,
EP-EFAs model reduces the complexity of EFAs by removing the state parameter,
which makes its opacity properties decidable.
We have provided the verification algorithms for the current-state opacity, initial-state opacity and
infinite-step opacity of EP-EFAs model, and discussed the complexity of these algorithms.

\par
One of the future work is to investigate the opacity enforcement of parametric DESs.
Another work worthy of further investigation is to explore a more powerful parametric  model
whose opacity properties are still decidable.

\section*{Acknowledgments}
This work is supported by the National Natural Science Foundation of China (Grant No. 61876195),
 the Natural Science Foundation of Guangdong Province of China (Grant No.  2022A1515011136),
 the Special Projects in Key Fields Foundation of the Department of Education of Guangdong Province of China
  (Grant No. 2021ZDZX1043), Guangxi Science and Technology Project (No. Guike AD23026227)
  and the Project Improving the Basic Scientific Research Ability of Young and Middle-aged Teachers
  in Guangxi Universities of China (Grant No. 2021KY0591).

\begin{IEEEbiography}[{\includegraphics[width=1in,height=1.25in,clip,keepaspectratio]{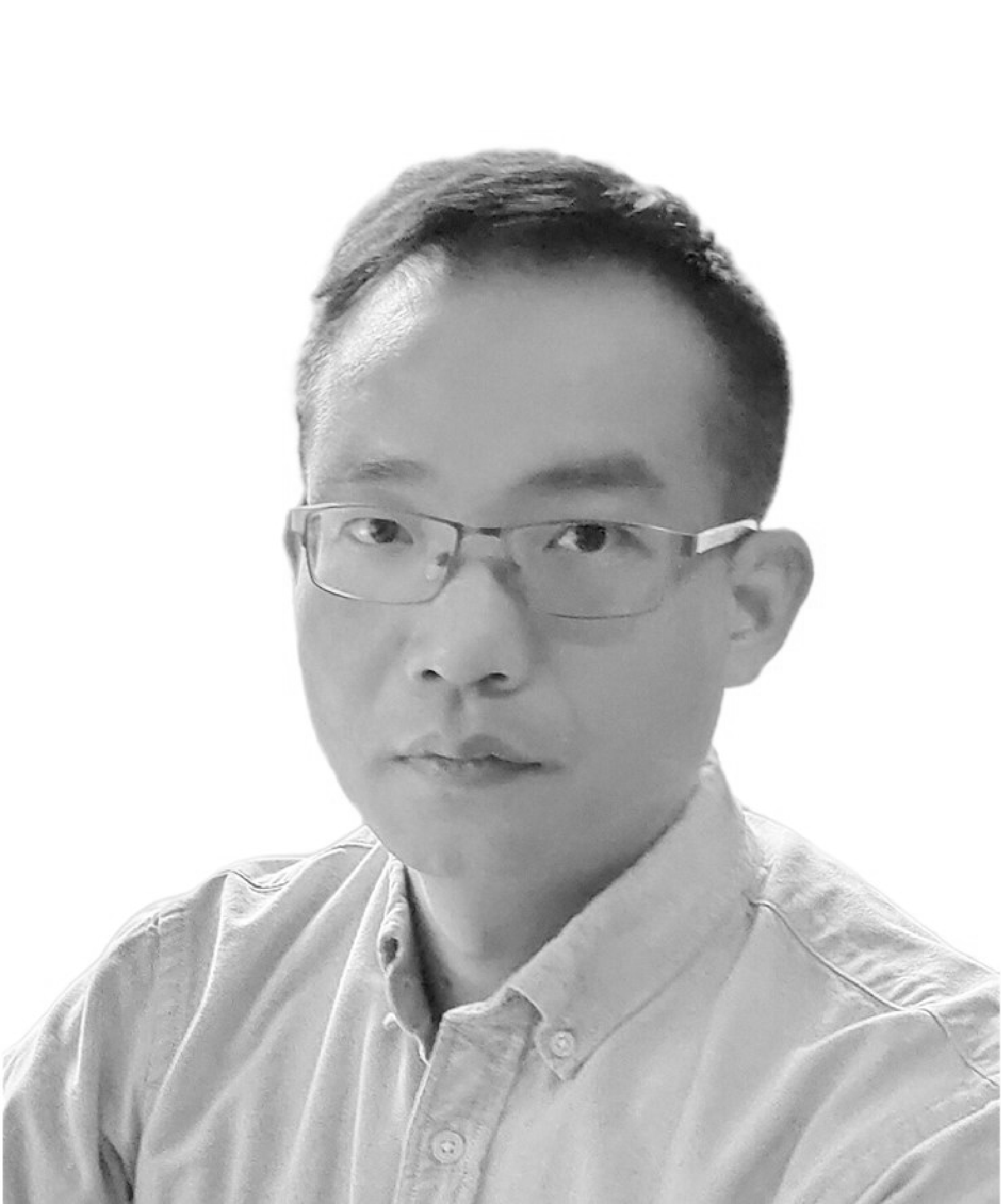}}]
{Weilin Deng}
 received the B.S. and M.S. degrees in computer science from
 South China University of Technology, Guangzhou, China, in 2003 and 2008, respectively,
 and the Ph.D. degree in computer software and theory from Sun Yat-Sen University, Guangzhou, China, in 2016.
 \par
 From 2016 to 2019, he was an associate research fellow with Sun Yat-Sen University.
 He is currently an associate professor with Guangdong University of Finance.
His current research interests include  discrete-event systems, fuzzy/probabilistic systems and computations,
and theoretical computer science. He is the author or co-author of more than 20 peer-review
papers published in various academic journals and conferences, including IEEE TAC, IEEE TFS, IEEE CDC,
INT J CONTROL and Information Sciences.
\end{IEEEbiography}

\begin{IEEEbiography}[{\includegraphics[width=1in,height=1.25in,clip,keepaspectratio]{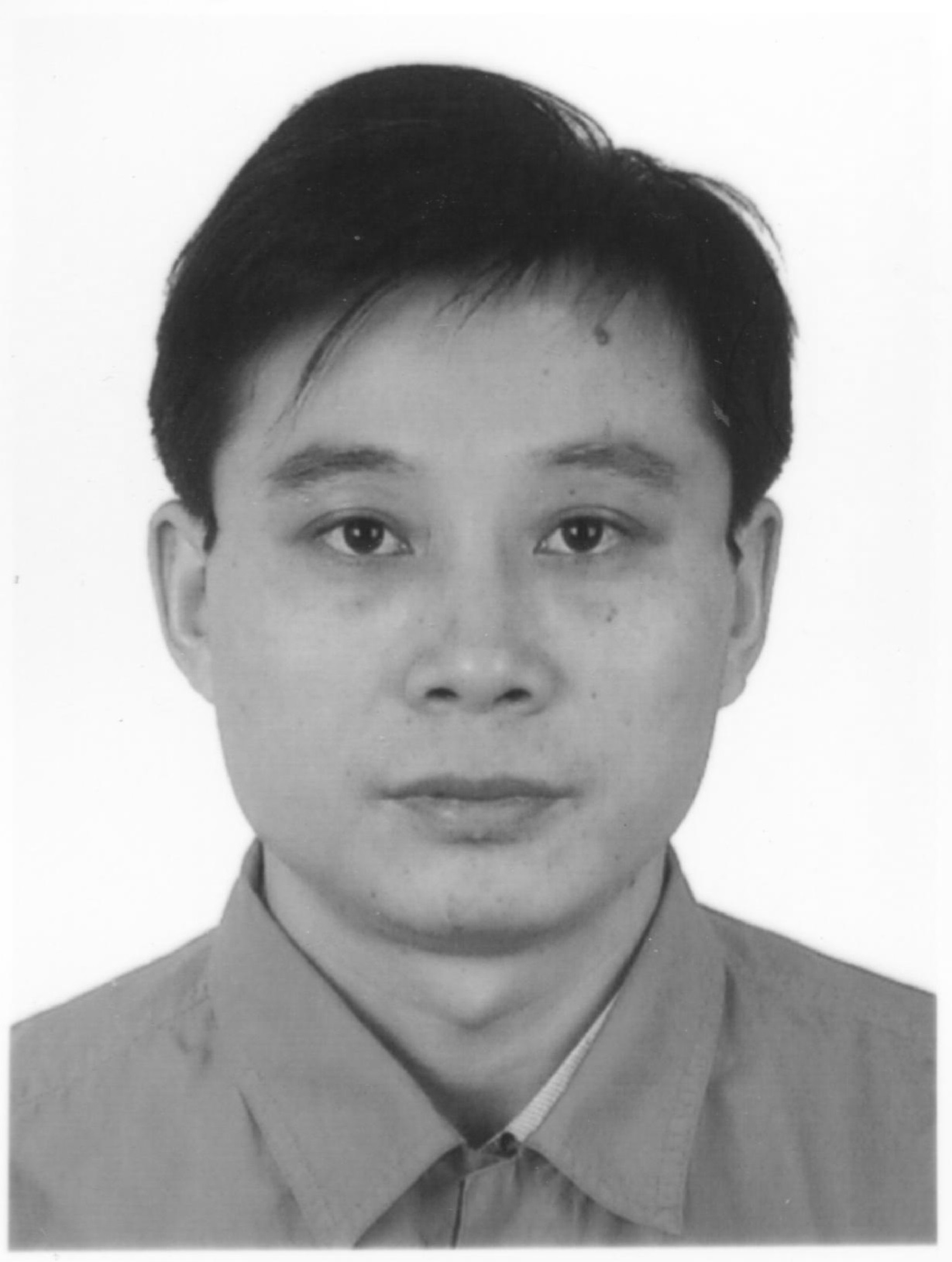}}]
{Daowen Qiu} received the M.S. degree in mathematics from Jiangxi Normal University, Nanchang, China, in 1993 and
the Ph.D. degree in mathematics from Sun Yat-Sen University, Guangzhou, China, in 2000.
\par
During 2000 and 2001, he was a Postdoctoral Researcher in computer science with Tsinghua University, Beijing, China.
Since August 2002, he has been associated with Sun Yat-Sen University, and then a Full Professor of computer science
 in May 2004. His current research interests include quantum computing, discrete-event systems, fuzzy and
probabilistic computation, and he has focused on models of quantum and probabilistic computation, quantum information.
He is the author or co-author of more than 160 peer-review papers published in various academic journals and conferences,
including Information and Computation, Artificial Intelligence, Journal of Computer and System Sciences,
Theoretical Computer Science, IEEE TRANSACTIONS ON SYSTEMS, MAN, AND CYBERNETICS PART B,
IEEE TRANSACTIONS ON AUTOMATIC CONTROL, IEEE TRANSACTIONS ON FUZZY SYSTEMS, Physical Review A,
Quantum Information and Computation, Journal of Physics A, and Science in China.
He is an editor of Theoretical Computer Science.
\end{IEEEbiography}

\begin{IEEEbiography}[{\includegraphics[width=1in,height=1.25in,clip,keepaspectratio]{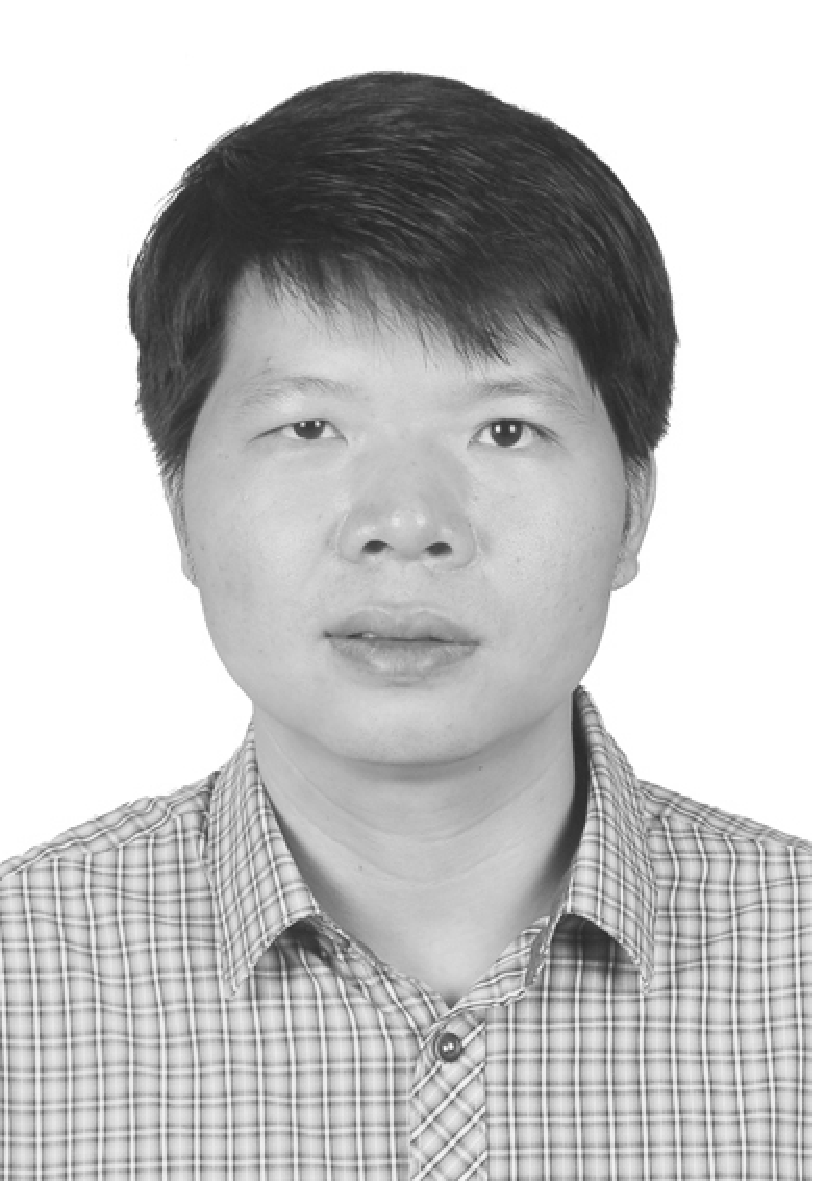}}]
{Jingkai Yang}
 received the B.S. and M.S. degrees in mathematics from  Guangxi Normal University, Guilin, China, in 2006 and 2009,
 respectively, and the Ph.D. degree in computer science and technology from Sun Yat-Sen University,
 Guangzhou, China, in 2022.
 He is currently an associate professor with Yulin Normal University.
 His main research interests include opacity analysis, supervisory control and failure diagnosis of
 discrete-event systems.
\end{IEEEbiography}

\end{document}